\documentstyle[aps,pre]{revtex}
\begin{document}
\draft
\def\be{\begin{equation}}
\def\ee{\end{equation}}
\def\bfi{\begin{figure}}
\def\efi{\end{figure}}
\def\bea{\begin{eqnarray}}
\def\eea{\end{eqnarray}}
\title{The overall time evolution in phase-ordering kinetics}
\author{Claudio Castellano\thanks{claudio@ictp.trieste.it}}
\address{The ``Abdus Salam'' I.C.T.P.,
Strada Costiera 11, I-34100 Trieste, Italy}
\author{Marco Zannetti\thanks{zannetti@na.infn.it}}
\address{Istituto Nazionale di Fisica della Materia, Unit\`{a}
di Salerno and Dipartimento di Fisica, Universit\`{a} di Salerno,
I-84081 Baronissi (Salerno), Italy}
\maketitle
\begin{abstract}
The phenomenology from the time of the quench to the
asymptotic behavior in the phase-ordering kinetics of a system with
conserved order parameter is investigated in the Bray-Humayun
model and in the Cahn-Hilliard-Cook model. From the comparison
of the structure factor in the two models the generic pattern
of the overall time evolution, based on the sequence ``early
linear - intermediate mean field - late asymptotic regime'' is
extracted. It is found that the time duration of each of these
regimes is strongly dependent on the wave vector and on the parameters
of the quench, such as the amplitude of the initial fluctuations
and the final equilibrium temperature. The rich and complex 
crossover phenomenology arising as these parameters are varied
can be accounted for in a simple way through the structure 
of the solution of the Bray-Humayun model.
\end{abstract}

\section{Introduction}
Phase-ordering kinetics is very much an open problem in 
non equilibrium statistical physics.
Its quite simple formulation is in striking contrast with the
difficulties encountered in its study, due to the strong
non linear nature of the problem.
During the past two decades much attention has been devoted to this
field and much progress has been made in the understanding of what
goes on after a quench inside the unstable region of the phase
diagram\cite{Bray94}.
In particular, the late stage of the evolution has been studied
with great care leading to a detailed knowledge of the scale 
invariant asymptotic time regime.
Despite this progress, however, most results recently obtained
are still based on uncontrolled {\it ad hoc} approximations relying on
the unproven dynamical scaling hypothesis, while a systematic theory
allowing for the derivation of late stage properties from first principles
is still far to come\cite{Mazenko88}.

Leaving aside this difficult problem, in this paper we mainly 
focus on what happens
before the scaling regime is entered, aiming at a comprehensive description
of the overall time evolution in a system with scalar conserved order
parameter.
Specifically, we will consider the Cahn-Hilliard-Cook (CHC) 
model\cite{Cahn58,Cook70} or model B in the classification of Hohenberg and
Halperin\cite{Halperin77}.
The study of the preasymptotic regime\cite{Binder84,Elder88}
has been hitherto somewhat neglected
because of the absence of scaling and universality.
This feature on one hand makes the understanding of the preasymptotics
less urgent and on the other makes it more difficult a subject to study
than the late stage.
As a result, all what we have in terms of early stage theory essentially
boils down to the linear approximation\cite{Cahn58}.
Nevertheless, it seems clear that an understanding of the complex
phenomena taking place in the transition from the preasymptotic to the
asymptotic regime is an important preliminary step in the development of a
full theory of phase-ordering kinetics\cite{Elder88,Langer92}.
In this respect, the effort undertaken in the last few 
years\cite{Coniglio89,Castellano97} to make an
assessment of the relevance of the large-$N$ model\cite{Mazenko84} 
in the theory of
phase-ordering kinetics has turned out to be quite productive.
Summarizing briefly, the large-$N$ model is the only exactly soluble
model presently available for phase-ordering kinetics with conserved
order parameter.
The asymptotic behavior of the exact solution for the structure factor in
this model displays multiscaling\cite{Coniglio89} 
in place of the standard scaling
usually observed in experiments and simulations.
Then, the question was raised whether multiscaling is the true asymptotic
behavior in all cases.
The issue was settled by Bray and Humayun (BH) by 
introducing\cite{Bray92,Castellano96} a
more general model (BH model) which contains the large-$N$ model as
a particular case and which displays standard scaling in the asymptotic
regime for all finite values of $N$, where $N$ is the number of  order
parameter components. From this result follows that multiscaling
as an asymptotic property holds
only if $N$ is strictly infinite, while for finite $N$ at most
it may be observable as a preasymptotic feature.

Continuing this line of reasoning, in this paper we carry out 
a detailed analysis of the BH model arriving to 
a fairly complete picture of the mechanism regulating
the structure of the preasymptotic behavior and, what is more
important, we find that it applies also to the CHC model.
The main result then is that what appears to be the generic
pattern in the overall time evolution is based on the sequence
"early linear - intermediate non linear large-$N$ (or mean field)
- asymptotic fully nonlinear behavior" {\it combined} 
with a strong dependence of the time scales of these regimes on the length
scale.
To be more specific, in wave vector space we identify three regions 
characterized by greatly different durations of the 
preasymptotic regimes and, furthermore,
these durations can be modulated by the choice of the parameters of the
quench. The three regions are the one around the peak of the structure factor
and the two to the right and to the left of the peak.
The concurrence of all these elements gives rise to a rich variety of
behaviors in the preasymptotic phenomenology which can be nicely
sorted out analytically on the basis of the equation for the BH model,
but holds true for the full CHC model as well.

The paper is organized as follows: in Section~\ref{SecModel} we introduce 
the model and after defining the observables of interest, we discuss
the linear, mean field and BH approximations.
The numerical solution of the BH model is analyzed in Section~\ref{SecBH}.
The corresponding analysis of the simulation of the CHC model is presented
in Section~\ref{SecCDS}.
Finally, Section~\ref{SecConc} contains concluding remarks.
A partial preliminary account of these results was published 
previously\cite{Castellano96b}.

\section{The model}
\label{SecModel}
We consider a system described by an $N$-component order parameter
which evolves according to the CHC equation
for conserved order parameter
\be
{\partial {\vec \phi}({\vec x},t) \over \partial t} =
\nabla^2 
\left[ {\partial V({\vec \phi}) \over \partial {\vec \phi}} 
- \nabla^2{\vec \phi} \right]
+ {\vec \eta}({\vec x},t)
\label{4.1}
\ee
where $V({\vec \phi}) = {r \over 2} {\vec \phi}^2 + {g \over 4N}
({\vec \phi}^2)^2$, with $r<0$ and $g>0$, is the mexican hat potential and
${\vec \eta}$ is a gaussian white noise, with expectations
\be
\left \{
\begin{array}{ccl}
<{\vec \eta}({\vec x},t)> & = & 0 \\
<\eta_{\alpha}({\vec x},t)\eta_{\beta}({\vec x}',t')> & =
& - 2 T_F  \delta_{\alpha \beta} \nabla^2
\delta({\vec x}-{\vec x}')\delta(t-t').
\end{array}
\right.
\ee
Here $T_F$ is the final temperature of the quench.
In the following we shall consider processes with an high temperature
uncorrelated initial state
\be
\left \{
\begin{array}{ccl}
<{\vec \phi}({\vec x},0)> & = & 0 \\
< \phi_{\alpha}({\vec x},0) \phi_{\beta}({\vec x}',0)> & = &
\Delta \delta_{\alpha \beta} \delta({\vec x}-{\vec x}')
\end{array}
\right.
\ee
where $\Delta$ is the size of the initial fluctuations.
In the subsequent time development the average order parameter remains zero
and the observable of interest is the equal time structure factor
$C({\vec k},t)$ given by the Fourier transform of the generic component
of the correlation function $<\phi_\alpha(\vec{x},t) \phi_\alpha(\vec{x}',t)>$.

If the final temperature is below the critical point we know that the
structure factor evolves from the initial form
\be
C(\vec{k},0)=\Delta
\label{C_in}
\ee
towards the equilibrium form
\be
\lim_{t \to \infty} C(\vec{k},t) = M^2(T_F) \delta({\vec k}) +
C^{(eq)}_{T_F}(\vec{k})
\label{C_eq}
\ee
where the first term is the Bragg peak associated to ordering, with an
equilibrium value $M(T_F)$ for the order parameter in the broken symmetry
phase, and $C^{(eq)}_{T_F}(\vec{k})$ is the contribution of thermal fluctuations.
The problem is to describe how this form of the structure factor evolves out
of the initial form~(\ref{C_in}).
Parameters of the quench are $\Delta$ and $T_F$.

The physical picture born out of experiments and simulations is clear:
after a certain initial transient, order is established over domains of size
$L(t)$ and within domains thermal fluctuations have relaxed to equilibrium.
Once this arrangement is achieved, memory of the initial condition is lost
and the subsequent time evolution (late stage) is essentially limited
to the coarsening of the ordered domains with thermal fluctuations
enslaved in the equilibrium form.
Phenomenological arguments\cite{Bray94} tell how 
the size of domains grows in time
$L(t) = k_m^{-1}(t) \sim t^{1/z}$, where $k_m(t)$ is the peak wave vector
of the structure factor and $z=3$ or $z=4$ in the scalar or vectorial case,
respectively.
At this stage, the structure factor can be split\cite{Mazenko88} 
into an ordering and
a fluctuation component which are, to a good approximation, decoupled
\be
C(\vec{k},t) = C_{ord}({\vec k},t) + C_{T_F}({\vec k},t). 
\label{C_tot}
\ee
The ordering component scales with respect to the size of domains
\be
C_{ord}({\vec k},t) = M^2(T_F) L^\alpha(x) F(x)
\label{C_SS}
\ee
with $\alpha=d$, the space dimensionality of the system, and $x=kL(t)$.
The above scaling form is a finite size representation of the Bragg peak and
reveals that the late stage evolution is a critical phenomenon characterized
by the pair of exponents $(\alpha,z)$.
The final temperature $T_F$ enters only in the prefactor $M^2(T_F)$
and the exponents are independent of both $\Delta$ and $T_F$.
In this sense the late stage is universal and scale invariant, with an
attractive fixed point at $T_F=0$.
Most of the recent theoretical work\cite{Bray94,Yeung94} 
has been concentrated on the study
of the scaling function $F(x)$, obtaining successful results through
approximations based on the assumption that scaling holds.
Much less is known about the process before the late stage is entered,
when the time evolution is still sensitive to the actual
values of the parameters ($\Delta$,$T_F$), 
the scaling form~(\ref{C_SS}) does not yet
hold and the separation into an ordering and a fluctuation component is a
much more delicate issue.
In the following we address the theoretical study of the preasymptotic
behavior.

\subsection{Linear theory}
The traditional early stage theory is the linear theory obtained by setting
$g=0$ in Eq.~(\ref{4.1}).
It is then straightforward to compute the structure factor
\be
C({\vec k},t) = \Delta C^{(l)}_0({\vec k},t) +
T_F C^{(l)}_{T_F}({\vec k},t)
\label{9.1}
\ee
where
\be
C^{(l)}_0({\vec k},t) = \exp \left[ -2k^2(k^2+r)t \right ]
\label{9.2}
\ee
and
\be
C^{(l)}_{T_F}({\vec k},t) = {1 \over k^2+r} \left \{ 1 -
\exp \left[ -2k^2(k^2+r)t \right ] \right \}.
\label{9.3}
\ee
The distinctive features of linear behavior are:
\begin{itemize}
\item
exponential growth for $k<k_0 \equiv \sqrt{-r}$.
In particular the peak wave vector of the structure factor remains at a
constant value $k_m= k_0/\sqrt{2}$ (except for very short times when $T_F>0$,
see Appendix) and
\be
C({\vec k_m},t) \sim \exp{2 k_m^4 t}
\label{C_m_lin}
\ee
\item
at $k=k_0$
\be
C({\vec k_0},t) =  
\left \{
\begin{array}{lcl}
\Delta        & \hspace{1cm} &\mbox{for } T_F=0 \\
2 T_F k^2_0 t & \hspace{1cm} &\mbox{for } T_F>0
\end{array}
\right.
\ee
\item
for $k>k_0$ the structure factor relaxes exponentially to 
\be
C({\vec k}) =  
\left \{
\begin{array}{lcl}
0             & \hspace{1cm} &\mbox{for } T_F=0 \\
T_F/(k^2+r)   & \hspace{1cm} &\mbox{for } T_F>0.
\end{array}
\right.
\label{ClinTF}
\ee
\end{itemize}
This approximation is useful to describe what happens in the very early
stage, when the requirement for the linear approximation to hold is met.
Namely, when the size of the nonlinear term in Eq.~(\ref{4.1}) is small
with respect to the linear one.
For the initial condition this implies $\Delta \ll M^2_0 = -r/g$ (small
$\Delta$), where $M_0$ is the equilibrium value of the order parameter at
zero temperature, at the bottom of the mexican hat potential.
Conversely, if $\Delta \simeq M^2_0$ (large $\Delta$), the linear theory
does never apply.
However, even when it applies at the beginning, at some point this
approximation
is bound to break down due to the exponential growth for $k<k_0$.
Furthermore, the form~(\ref{9.1}) does not describe coarsening,
since the peak value of the wave vector remains constant in
time, and the scaling form~(\ref{C_SS}) never applies.
In short, the linear early stage theory does not connect to the late
stage theory.
This is true also to any finite order in $g$.
Therefore, any approximation aiming at a global description of the
phase-ordering process must necessarily involve some form of
infinite resummation of the expansion in powers of $g$.

\subsection{Large-$N$ model}
By analogy with the theory of critical phenomena, one may explore the
$1/N$ expansion as a method to organize infinite partial resummations.
To lowest order (large-$N$ model), one obtains from~(\ref{4.1})
the following equation for the structure factor
\be
{\partial C({\vec k},t) \over \partial t} = 
-2 k^2  [k^2+R(t)] C({\vec k},t) + 2 k^2 T_F 
\label{C}
\ee
with
\be
R(t)= r + g S(t)
\label{R}
\ee
and
\be
S(t) = \int {d^dk \over (2\pi)^d} C({\vec k},t). 
\label{S}
\ee
This equation can be solved analytically\cite{Coniglio89} 
yielding the full time evolution
from the initial condition~(\ref{C_in}) to the final equilibrium
form~(\ref{C_eq}).
However, the time development of $C({\vec k},t)$ obtained in the late stage
of this model differs in several respects from the expected
behavior~(\ref{C_tot}) and~(\ref{C_SS}).
First of all, the contribution of the structure factor associated to the
buildup of the Bragg peak takes the multiscaling form
\be
C({\vec k},t) \sim [L(k_{m}L)^{2/d-1}]^{\alpha(x)} F(x)
\label{C_MS}
\ee
where $L(t)=t^{1/4}$, $k_m(t)$ is the peak wave vector related to $L(t)$ by
\be
(k_m L)^4= d \log L +(2-d) \log(k_m L)
\label{k_mlog}
\ee
and $x=k/k_m(t)$.
Then, the exponent $\alpha(x)$ and the scaling function $F(x)$ do take
different forms for the quenches to finite temperature with $0<T_F<T_c$
and for $T_F=0$.
In the former case
\be
\alpha(x) = \alpha_{T}(x) \equiv
\left \{
\begin{array}{lcl}
2 + (d-2) \psi(x) & \hspace{1cm} & x < x^* \\
2                 & \hspace{1cm} & x > x^*
\end{array}
\label{MST}
\right.
\ee
and
\be
F(x)= {T_F \over x^2}
\ee
with $x^* = \sqrt{2}$ and $\psi(x) = 1 - (1-x^2)^2$, while in the
latter
\be
\alpha(x) = \alpha_{0}(x) \equiv d \psi(x)
\label{MS0}
\ee
and
\be
F(x)=1.
\ee

This result shows that the infinite resummation involved in the lowest order
of the $1/N$ expansion, although producing equilibration and formation
of the Bragg peak, it does so in a qualitatively different way from what
expected in the full CHC theory, since i) the finite size 
representation~(\ref{C_MS}) of the Bragg 
peak displays multiscaling in place of the
standard form of scaling~(\ref{C_SS}) and ii) the fixed point structure
regulating the asymptotic scaling properties is different as the spectrum
of exponents and the scaling function are not anymore independent of the 
final value of the temperature.
At $T_F=0$, in place of an attractive fixed point, 
there is an isolated fixed point. Then, there is a line of fixed points for
$0<T_F<T_c$ (all sharing the same asymptotic properties) 
and, finally, there is yet another isolated fixed point at $T_c$.
It has now been clarified\cite{Castellano97} that the formation of the Bragg peak in the
large-$N$ model is associated to a condensation process rather than to the
phase-ordering process.
Why this ought to induce multiscaling is not clear.
In any case, this seems to say that the type of resummation involved
in the large-$N$ model is not enough to produce phase-ordering, neither
phase-ordering can be recovered by perturbation theory about the
$N=\infty$ limit.
In the following we shall refer to the behavior described above as mean field
behavior.

In order to expose some features of the solution which are relevant
also beyond mean field theory, it is convenient to consider the integral
form of Eq.~(\ref{C})
\be
C({\vec k},t) = \Delta C^{(mf)}_0({\vec k},t) +
T_F C^{(mf)}_{T_F}({\vec k},t)
\label{C_mf_tot}
\ee
which yields the structure factor as the sum of two contributions
\be
C^{(mf)}_0({\vec k},t) = \exp{\left \{-2 k^2\int_0^t dt' [k^2+R(t')]\right \}}
\label{C_mf_0}
\ee
\be
C^{(mf)}_{T_F}({\vec k},t) = 2 k^2 \int_0^t dt' {C^{(mf)}_0({\vec k},t) \over
C^{(mf)}_0({\vec k},t')}
\ee
coupled together through the self-consistency relation required by the
definition of $R(t)$ in Eqs.~(\ref{R}) and~(\ref{S}).
If $T_F=0$ obviously only $C^{(mf)}_0({\vec k},t)$ enters the self-consistency
relation and this yields the asymptotic multiscaling behavior characterized
by $\alpha_0(x)$.
Instead, if $T_F \ne 0$, both terms in the right hand side
of~(\ref{C_mf_tot}) participate in the self-consistency relation.
Eventually $C^{(mf)}_{T_F}({\vec k},t)$ prevails and the asymptotic
multiscaling behavior~(\ref{C_MS}) with $\alpha_T(x)$ is obtained.
However, before reaching this regime, there may exist a time interval
where the two terms do compete.
This clearly depends on the relative magnitude of $\Delta$ and $T_F$.
Less obvious is that it should depend also on the length scale and that
there should be, therefore, a wave vector dependent crossover time $t^*(k)$.
In order to understand how this comes about, a glance at Fig.~(\ref{Figold})
is sufficient.
The exponent $\alpha(x)$ is related to the rate of growth of the structure
factor at some length scale.
Roughly, $\alpha_0(x)$ and $\alpha_T(x)$ indicate how fast, respectively,
$C^{(mf)}_0({\vec k},t)$ and $C^{(mf)}_{T_F}({\vec k},t)$ grow at the
different wave vectors as time goes on.
Fig.~(\ref{Figold}) shows that $\alpha_T(x)>\alpha_0(x)$ everywhere,
except at $x=1$, where $\alpha_0(1)=\alpha_T(1)=d$.
Hence, if $\Delta/T_F$ is sufficiently large for the behavior of
$C^{(mf)}_0({\vec k},t)$ to be observable at the beginning of the quench,
this will yield to the true asymptotic behavior of
$C^{(mf)}_{T_F}({\vec k},t)$ first where the difference
\be
\delta \alpha(x) = \alpha_T(x)-\alpha_0(x)
\ee
is the largest, and then gradually the crossover from
$C^{(mf)}_0({\vec k},t)$ to $C^{(mf)}_{T_F}({\vec k},t)$ will propagate in
time to the length scales where $\delta \alpha(x)$ is smaller.
Therefore, $\delta \alpha(x)$ is the key quantity which controls the
wave vector dependence of $t^*(k)$.
In the case at hand, the crossover will take place very fast for large values
of $x$ and will then propagate towards $x^*$ from above.
When it takes place around $x^*$ it occurs also for small values of
$x$, eventually propagating towards $x=1$ both from the right and the left.
Thus the peak of the structure factor is the wave vector region where the
crossover time is the longest.
The rich variety of behaviors generated by Eq.~(\ref{C_mf_tot}) as $\Delta$
and $T_F$ are varied has been studied in great detail in
Ref.~\cite{Castellano97}.

\subsection{Bray-Humayun model}
A significant improvement over the large-$N$ model is obtained in the
BH model.
By combining the gaussian auxiliary field approximation of 
Mazenko\cite{Mazenko89} with
the $1/N$-expansion, a non linear closed equation for the structure
factor is derived\cite{Bray92}
\begin{eqnarray}
{\partial C({\vec k},t) \over \partial t} & = &
-2 k^2  [k^2+R(t)] C({\vec k},t) \nonumber \\
 & & - 2{k^2 \over N} R(t) D({\vec k},t) + 2 k^2 T_F
\label{BH}
\end{eqnarray}
with
\be
D({\vec k},t) = \int {d^dk_1 \over (2\pi)^d } \int {d^dk_2 \over (2\pi)^d }
C({\vec k}-{\vec k_1},t) C({\vec k_1}-{\vec k_2},t) C({\vec k_2},t)
\ee
and $R(t)$ defined by~(\ref{R}) and~(\ref{S}).
Even if the large-$N$ model is contained in Eq.~(\ref{BH}) as a particular
case recovered in the limit $N \to \infty$, it is not possible to pinpoint
the correction made, due to the uncontrolled character of the approximation
involved in the gaussian auxiliary field method\cite{Yeung94}.
Neither it is clear how to proceed, at least in principle, in order to
improve over Eq.~(\ref{BH}).
In any case, the merit of the BH equation is that in the late stage
standard scaling is recovered for any finite value of $N$.
This means that, whatever correction is contained in the BH model,
it is enough to describe phase-ordering rather than the condensation
process appearing in the large-$N$ model.
Therefore the BH equation may be regarded as the self-contained definition
of a basic model for phase-ordering kinetics.
Even though the model does not allow for a complete explicit solution, the
structure of the formal solution in conjunction with numerical
analysis allows to follow in great detail the overall time development and
to uncover how the asymptotic scaling behavior is gradually born out
of the non universal preasymptotic behavior.

Going over to the integral form of Eq.~(\ref{BH}) we find the generalization
of Eq.~(\ref{C_mf_tot})
\be
C({\vec k},t) = \Delta C^{(BH)}_0({\vec k},t) +
T_F C^{(BH)}_{T_F}({\vec k},t) + {1 \over N} C^{(BH)}_{nl}({\vec k},t)
\label{C_BH_tot}
\ee
with
\be
C^{(BH)}_0({\vec k},t) = \exp{ \left \{-2 k^2\int_0^t dt' [k^2+R(t')]\right \}}
\ee
\be
C^{(BH)}_{T_F}({\vec k},t) = 2 k^2 \int_0^t dt' {C^{(BH)}_0({\vec k},t) \over
C^{(BH)}_0({\vec k},t')}
\ee
\be
C^{(BH)}_{nl}({\vec k},t) = -2 k^2 \int_0^t dt' {C^{(BH)}_0({\vec k},t) \over
C^{(BH)}_0({\vec k},t')} R(t') D({\vec k},t')
\ee
and where again the self-consistency is implemented through the definition
of $R(t)$.
In general this is a very complicated nonlinear integral equation which must
be handled numerically.
On the other hand, considerations of the type made about Eq.~(\ref{C_mf_tot})
may be extended to the present case and simplifications may be expected
if the conditions for one of the terms on the right hand side to prevail
over the others are realized.
First of all, the non universal features dependent on the parameters of
the quench ($\Delta$, $T_F$) are associated to the first two terms,
while the asymptotic behavior is due to the third one.
Taking only this last term into account, BH have been able to extract the
scaling behavior which obeys the standard form~(\ref{C_SS})
with 
\be
F(x) \sim  
\left \{
\begin{array}{lcl}
x^2               & \hspace{1cm} & \mbox{for } x \ll 1 \\
e^{-c_1 (x^2-1)^2 \ln N}   & \hspace{1cm} & \mbox{for } x \simeq 1 \\
e^{-c_2 x}        & \hspace{1cm} & \mbox{for } x \gg 1 
\end{array}
\right.
\label{F}
\ee
where $c_1$ and $c_2$ are constants. The exponential decay for large
$x$ has been obtained numerically\cite{Rao94} while the other 
two results are analytical\cite{Bray92,Rojas95}.
Before the asymptotics are reached, the behavior of the structure factor is
the result of the interplay of all the terms in the right hand side
of~(\ref{C_BH_tot}).
If there exists a time interval where the first two terms do actually
dominate over the third one, one may anticipate the occurrence of mean field
behavior\cite{Bray92,Castellano96}.
Furthermore, on the basis of the previous discussion, one may also anticipate
that this will be very much sensitive to the relative magnitude of $\Delta$,
$T_F$, $1/N$ and that there will be a wave vector dependent crossover time.

\subsection{Multiscaling analysis}

In order to cover all possible behaviors it is quite useful to introduce the
multiscaling analysis of the structure factor.
Both the standard scaling~(\ref{C_SS}) and the multiscaling~(\ref{C_MS})
behaviors can be written in a unified general form as
\be
C({\vec k},t) = \left [{\cal L}_1(t) \right]^{\alpha(x)} F(x)
\label{MSA}
\ee
with $x=k{\cal L}_2(t)$ and where ${\cal L}_1(t)$ and ${\cal L}_2(t)$ are two
growing lengths.
As we shall see, the determination of the spectrum of exponents $\alpha(x)$
is quite informative.
Numerically this is done by considering $C({\vec k},t)$ over subsequent
and non overlapping time intervals $\tau_i$.
After taking the logarithm of Eq.~(\ref{MSA}) one has
\be
\log C(x/{\cal L}_2(t),t)=\alpha(x,t) \log {\cal L}_1(t) + \log F(x)
\label{logMS}
\ee
and $\alpha(x,t)$ can be computed as the slope of 
$\log C({\vec k},t)$ versus $\log {\cal L}_1(t)$
for the different times in the interval $\tau_i$, with $x$ fixed and
$k=x/{\cal L}_2(t)$.
In practice this amounts to the measurement of an effective exponent
$\alpha(x,t) = {\partial \log C(k,t) \over \partial \log {\cal L}_1(t)}$
of the type introduced in critical phenomena for the study of crossovers
between competing critical behaviors.
Here the time $t$ is the parameter driving to criticality and
scaling holds when the effective exponent is independent of $t$.
Standard scaling corresponds to ${\cal L}_1(t)= {\cal L}_2(t) = L(t)
\equiv t^{1/z}$ and $\alpha(x)=d$, while
multiscaling corresponds to ${\cal L}_1(t)=L(t)[k_m(t)L(t)]^{2/d-1}$,
${\cal L}_2(t)=k_m^{-1}(t)$ and $\alpha(x)$ genuinely dependent on $x$.

>From the structure factor other quantities can be computed numerically
giving insight into the preasymptotic behavior.
In the following,
we will consider the position of the peak $k_m(t)$ and its height
$C(k_m,t)$.
Their asymptotic behaviors are
\be
k_m(t) \sim t^{-1/3}
\hspace{2cm}
C(k_m,t) \sim t^{d/3}
\ee
for the CHC model, while in the large-$N$ model from~(\ref{C_MS})
and~(\ref{k_mlog}) follows
\be
k_m(t) = \left({\log t \over t} \right)^{1/4}
\hspace{2cm}
C(k_m,t) = t^{d/4} (\log t)^{(2-d)/4d}.
\ee
 
For future reference it is useful to apply the multiscaling analysis
to the linear approximation.
When $T_F=0$ and $\Delta > 0$ from Eq.(\ref{9.2}) one finds
\be
\log C(xk_m,t) = {2 k_m^4 t \over \log \left[L (k_m L)^{2/d-1}\right]}
\psi(x) \log \left[L (k_m L)^{2/d-1}\right] + \log \Delta
\ee
which gives the effective exponent
\be
\alpha(x,t) = {2 k_m^4 t \over \log[L (k_m L)^{2/d-1}]} \psi(x)
\label{40}
\ee
namely the spectrum is of the multiscaling form~(\ref{MS0}) with a prefactor
linearly growing with time.
This explicit time dependence shows the absence of scaling.
For $\Delta=0$ and $T_F>0$
the computation of $\alpha(x,t)$ is slightly more complicated and
is reported in the Appendix. 

\subsection{Numerical solution}
The numerical solution of the CHC
equation~(\ref{4.1}) requires a lot of computer resources.
This is due to the need of reaching long times, in order
to study the asymptotic behavior, and of considering large systems,
in order to avoid finite size effects.
The task of analyzing in detail the effective exponent $\alpha(x,t)$
makes things worse because it crucially depends on a very precise
estimate of the structure factor. This involves the use of very
large systems and the average over many realizations of the initial
fluctuations and of the thermal noise.

 The Bray-Humayun model, since no averaging over realizations is
required, has been solved by simple discretization of the equation of motion
for the correlation function in real space, which
involves only laplacians. The size of the system has been chosen 
$1024^2$ for $d=2$ and $160^3$ for $d=3$.

For the scalar case we have resorted, instead, to the Cell 
Dynamical System (CDS),
a cellular automaton which reproduces accurately the solution
of the CHC equation, while being computationally much more
efficient\cite{Oono87}.
We have considered the version of the CDS algorithm with $f(x)=A \tanh(x)$ 
and have taken the parameters $A=1.3$ and $D=0.5$, so that
$M(T_F=0)=\pm 0.97767$.
We have considered lattices of size $512^2$ for $d=2$ and $128^3$ for $d=3$.
In the initial state we have considered a uniform distribution of
the order parameter between $-b/2$ and $b/2$ so that $\Delta = b^2/12$.
For each value of the parameters $\Delta$ and $T_F$ the
structure factor has been averaged over a suitable number of realizations.
The values of $k_m(t)$, $C(k_m,t)$ and $\alpha(x,t)$ have been obtained by
interpolating the spherically averaged value of $C({\vec k},t)$ with a
cubic spline routine.
 
\section{Results for the BH model}
\label{SecBH} 
In this section we present the multiscaling analysis of the structure
factor in the BH model.
As pointed out above, in the asymptotic regime scaling is of the standard type
as long as $1/N$ is not zero, because the non linear contribution
$C^{(BH)}_{nl}({\vec k},t)$ in Eq.~(\ref{C_BH_tot}) is eventually the largest
for all values of $x$.
Here we are mainly interested in the numerical study of the
preasymptotic regime and in the identification of patterns due to the
interplay of the three terms on the right hand side of Eq.~(\ref{C_BH_tot}).

\subsection{$T_F=0$}
We start by considering the case of a quench to zero final temperature, for
a system with $N=10^3$ and $\Delta= 10^{-8}$.
The overall evolution of the structure factor is rather complicate and
in order to sort out all the features entering it, very good quality of data
is needed.
Therefore, we present the data obtained for $d=2$ which are much less noisy
and are qualitatively the same as those for $d=3$, since $T_F=0$.
With $T_F=0$ the competition is reduced to the first and the third term
in the right hand side of Eq.~(\ref{C_BH_tot}).
Following the reasoning presented in Section~\ref{SecModel}, this depends
on the values of $\Delta$ and $N$ and on the rates of growth of the two
terms at the different wave vectors.
Fig.~(\ref{Fig01}) is a double logarithmic plot of the time evolution of
$C({\vec k},t)$ from the very beginning up to some time inside the
asymptotic regime.
Basic features are the prompt appearance of power law tails on both sides
of the peak and a more elaborate behavior of the peak.
In particular, three sharp regimes are identified 
(early, intermediate and late stage) which are
clearly separated by almost abrupt changes in the position $k_m(t)$ and
height $C(k_m,t)$ of the peak at times $t_1$ and $t_2$ (see also
Figs.~(\ref{Fig03}) and~(\ref{Fig04})).
In the early stage, i.e. in the interval (0,$t_1$), there is no coarsening
and $k_m$ stays constant in time with high accuracy (Fig.~(\ref{Fig03}))
as predicted by the linear approximation.
That there should be a linear regime could have been anticipated from the
tiny value of the initial fluctuations and it is further
confirmed by the behavior of $C(k_m,t)$ in Fig.~(\ref{Fig04}), which
obeys well Eq.~(\ref{C_m_lin}).
On the sides of the peak, according to the linear approximation~(\ref{9.2}),
there should be an exponential decay in $k$, both to the right
$C^{(l)}_0(k,t) \sim e^{-2 k^4 t}$ and to the left
$C^{(l)}_0(k,t) \sim e^{-2 k^2 r t}$.
This holds true immediately after the quench, but very soon (i.e. for 
$t \ll t_1$), even if the peak keeps on staying in the same place,
the exponential decays are replaced by power laws.
These are obviously nonlinear effects.
The competition between $C^{(BH)}_0({\vec k},t)$ and
$C^{(BH)}_{nl}({\vec k},t)$ leading to the breakdown of the linear
approximation follows a non trivial pattern with $C^{(BH)}_{nl}({\vec k},t)$
taking over first at small and large wave vectors and then closing towards
the peak both from the right and the left.
The propagation of the breakdown of the linear approximation from
high to low wave vectors has already been discovered and analyzed in
detail in the non conserved case\cite{Gross94}.
Here, we find evidence that, through a different mechanism, this breakdown
follows a yet more rich pattern characterized by the bilateral convergence
towards the peak.

For what concerns the nature of the power laws, for small wave vectors
we find $C(k,t) \sim k^2$ implying that the asymptotic behavior~(\ref{F})
is established almost from the start.
For large wave vectors, instead, we find $C(k,t) \sim k^{-n}$ with $n \sim 4$.
Interestingly, even though this non linear feature sets in even before
the appearance of the $k^2$ power law for small $k$, it is very different
from the exponential decay of Eq.~(\ref{F}) and it lasts quite long before
yielding  to the formation of the true asymptotic behavior.
This phenomenon is not related to the interplay of terms in the right
hand side of~(\ref{C_BH_tot}). 
Rather, it may be related to the formation in the early
stage of unstable localized topological defects.
Such phenomenon is present also in simulations of the CHC equation with
$n>d$ and will be studied elsewhere\cite{Castellano98}.
In any case, the bottom line is that on the tails there is practically
no competition between $C^{(BH)}_0({\vec k},t)$ and
$C^{(BH)}_{nl}({\vec k},t)$, as the latter contribution soon prevails,
while a long lasting competition takes place around the peak.
We will thus concentrate on this region.

Figs.~(\ref{Fig03}) and~(\ref{Fig04}) show that immediately after
the end of the early regime, at the time $t_1$, coarsening begins with
power law behaviors for $k_m \sim t^{1/z}$ and $C(k_m,t) \sim t^{d/z}$.
However, after some time, both these quantities 
display a pronounced anomaly localized around
$t_2$ whose origin seems to be puzzling.
This is especially true considering the behavior of $k_m(t)$, which obeys the
same power law very precisely, with $z=4$, before and after 
the anomaly (see Fig.~(\ref{Fig03})).
For the peak height, while the power law is well obeyed with
$d/z=0.53$ for $t>t_2$, the behavior in the intermediate regime $t_1<t<t_2$
cannot be easily fitted to a power law.
The insight into what is going on is obtained by making a data collapse
of the structure factor (Fig.~(\ref{Fig02}))
which gives evidence for the existence of a crossover, 
in the region around the peak, between two
different scaling forms. Both the preasymptotic and the asymptotic
scaling functions are well fitted by the form
\be
F(x) \sim e^{-a(x^{2}-1)^{2}}
\label{scalingfunction}
\ee
which describes the large-$N$\cite{Coniglio89} and the BH structure
factor\cite{Bray92,Rojas95} around the peak. The change in the value
of $a$ in going from one regime to the other means that around the peak
$C^{(BH)}_0({\vec k},t)$ prevails over $C^{(BH)}_{nl}({\vec k},t)$ 
long enough not only to go past the linear region, but also to leave behind
the intermediate time interval ($t_1$,$t_2$) where the dynamics is non linear
with the self-consistency relation in Eq.~(\ref{C_BH_tot}) saturated to a good
extent  by $C^{(BH)}_0({\vec k},t)$ alone.
As a result, non linear mean field behavior shows up around the peak.

The picture of the overall evolution is completed by the behavior
of the effective exponent $\alpha(x,t)$.
In Fig.~(\ref{Fig00}), referring to times within the early stage,
the mechanism of breakdown of the linear approximation described above is
clearly displayed.
For all the curves, except the last one, there are two qualitatively different
behaviors for $x>{\hat x}(t)$ and  $x<{\hat x}(t)$, with ${\hat x}(t)$
decreasing with time.
For $x<{\hat x}(t)$ the effective exponent $\alpha(x,t)$ is well described
by the linear approximation~(\ref{40}), while for $x>{\hat x}(t)$ is
essentially flat.
This corresponds to the onset of the power law to the right of the peak in
the structure factor in Fig.~(\ref{Fig01}).
The retarded appearance  of the other power law tail to the left of the
peak corresponds to the deviation from the linear form of $\alpha(x,t)$
for small $x$ in the last curve of Fig.~(\ref{Fig00}).
The timings of the deviations from linear behavior at the different
values of $x$ are a consequence of the shape~(\ref{40}) of the 
effective exponent. Breakdown of linear behavior shows up first
at large $x$, where $\alpha(x,t)$ is large and {\it negative} and
later on around $x=0$ where $\alpha(x,t)$, although small, is
positive. 

Next, the behavior of $\alpha(x,t)$  over the 
subsequent time history in Fig.~(\ref{Fig05})
shows the formation of very pronounced peaks to the right and to the left of
$x=1$.
These are due to the very fast growth of the power law tails in the early
stage, as it is evident from the spacing of the curves in Fig.~(\ref{Fig01}).
In this time regime the peak is the slowest growing mode in the system.
It should be noted the multiscaling behavior around $x=1$, 
which is produced by the mean field behavior 
of the peak in the intermediate time regime.
Later on the curves for $\alpha(x,t)$ flatten and collapse around the constant
value $\alpha(x)=2$.
The disappearance of the time dependence from $\alpha(x,t)$ reflects the onset
of standard scaling in the late stage.
It should be noted, however, that the approach to asymptotic scaling is quite
a slow process, since a small upward curvature of $\alpha(x)$ persists
for a very long time.
This curvature was observed in early measurements of $\alpha(x)$ in a variety
of systems and was correctly interpreted as evidence that the large time
behavior, although not yet fully asymptotic, was definitely approaching
the standard scaling behavior\cite{Coniglio92}.

As a test of the above analysis let us consider the expected behavior upon
making large ($\Delta=1$) the initial fluctuations and smaller $1/N
(N=10^{6})$.
The immediate consequences should then be the total absence of a linear
regime and a considerable extension of the time interval where
$C^{(BH)}_0({\vec k},t)$ prevails over $C^{(BH)}_{nl}({\vec k},t)$.
It turns out that in this case it is not crucial to have extremely
precise data, so we present results for $d=3$.
Inspection of the evolution of the structure factor (Fig.~(\ref{Fig06}))
reveals a different morphology from the case of small $\Delta$ and the
absence of the linear behavior of the peak is quite evident.
Furthermore, the behavior of the tails is less dramatic and a power law
appears only at the largest times for large values of the wave vector.
More precisely, looking at the time dependence of $k_m(t)$ and $C(k_m,t)$ in
Figs.~(\ref{Fig03}) and~(\ref{Fig04}) we can recognize an early regime,
which lasts up to a time very close to $t_1$, followed by a second regime
characterized by power laws.
The features of the early regime can be explained from the behavior
of $C^{(BH)}_0({\vec k},t)$ for short time.
Namely, the exponential in the solution of the large-$N$
model~(\ref{C_mf_0}) is peaked at
\be
k_m^2 = - {\int_0^t dt' R(t') \over 2t}.
\ee
For very small $t$ one has $R(t') \simeq r+gS(0)$ and therefore
$k_m^2 \simeq -(r+gS(0))/2$.
Since in the case considered $r+gS(0)>0$ we have $k_m^2<0$. This means
that for very short times $C(k,t)$ is monotonically decreasing and the
maximum is at $k=0$.
During this short time interval (not shown in our plots) $k_m^{-1}= \infty$
and $C(k_m,t)=\Delta$.
As time proceeds, $S(t)$ gets smaller and rapidly
$\int_0^t dt' R(t')$ becomes negative while $k_m^2$ becomes positive
and starts growing.
This time interval corresponds to the decreasing behavior of $k_m^{-1}(t)$ in
Fig.~(\ref{Fig03}) for $\log(t)<0$.
Therefore, in the early regime the right hand side
of Eq.~(\ref{C_BH_tot}) is 
dominated by the contribution $C^{(BH)}_0({\vec k},t)$.

Later on coarsening begins with power laws both for $k_m(t)$ and $C(k_m,t)$.
The behavior of the effective exponent in Fig.~(\ref{Fig08})
shows scaling since $\alpha(x,t)$ becomes time independent.
The shape of $\alpha(x)$ obeys quite well the mean field multiscaling
form~(\ref{MS0}) for $x<x^*$ and the standard scaling form $\alpha(x)=d=3$
for $x>x^*$.
Namely, in the region where $\alpha_0(x)$ is negative
$C^{(BH)}_{nl}({\vec k},t)$ dominates, while in the region where
$\alpha_0(x)$ is positive, even though $\delta \alpha(x) = d - \alpha_0(x)$
is also positive, due to the large value of $\Delta$ it is 
$C^{(BH)}_0({\vec k},t)$ that prevails, and not just around the peak,
but for all values $x<x^*$.
This is an intermediate scaling regime, characterized by mean field
behavior over an extended range of wave vectors.
Eventually, there must be the crossover to the asymptotic regime
with standard scaling over all length scales.
This occurs for times larger than those considered in our computation.

\subsection{$T_F>0$}
With a finite final temperature all three terms are present in the right
hand side of Eq.~(\ref{C_BH_tot}).
Without going to the most general case, for our purposes it is
sufficiently illustrative to consider $T_F \gg 1/N$.
In this case, we expect to find, after the early stage, a fairly long
intermediate
regime characterized by the finite temperature mean field behavior
discussed in Section~\ref{SecModel}.
The eventual asymptotic regime lies beyond the time interval considered
in the computation.
In the opposite limit $T_F \ll 1/N$, the intermediate regime is expected
to be very much similar to the one analyzed above with $T_F=0$.

Choosing $d=3$, $N=10^6$ and $T_F=10^{-3}$ let us first consider the behavior
of the effective exponent $\alpha(x,t)$ when $\Delta=0$.
In this case an early linear regime is expected.
This is demonstrated by the set of the first few curves in Fig.~(\ref{Fig09}),
which follow closely the behavior of Eq.~(\ref{12.4}) in the Appendix. 
In particular,
the signature of linear behavior with finite temperature is given by the
relaxation of $\alpha(x,t)$ to zero for $x>x^*$, which corresponds to the
fast relaxation~(\ref{ClinTF}) to the time independent value of $C({\vec k},t)$
for $k>k_0$.
After the early stage, Fig.~(\ref{Fig09}) displays the onset of the
multiscaling regime with a shape of $\alpha(x)$ well represented by
$\alpha_T(x)$ in Eq.~(\ref{MST}).
Hence, with this choice of the parameters, in the intermediate regime
$C^{(BH)}_{T_F}({\vec k},t)$ dominates the right hand side of
Eq.~(\ref{C_BH_tot}), yielding finite temperature mean field behavior.
Changing the size of the initial fluctuations to the large value $\Delta=1$
(Fig.~(\ref{Fig10})), in the intermediate regime we find a different
multiscaling behavior.
For $x<x^*$, $\alpha(x)$ is well represented by $\alpha_0(x)= 3 \psi(x)$,
while for $x>x^*$ $\alpha(x) \simeq 2$ suggesting that in this region
$\alpha(x)=\alpha_T(x)$.
Hence, we find a behavior of the type described in the discussion of the
large-$N$ model in Section~\ref{SecModel}.
Namely, the first two terms dominate the third one in
Eq.~(\ref{C_BH_tot}), and among them $C^{(BH)}_0({\vec k},t)$ dominates
for $x<x^*$, while $C^{(BH)}_{T_F}({\vec k},t)$ dominates for $x>x^*$.
The limitation of CPU time has not allowed to pursue the time evolution any
further.
However, on the basis of Eq.~(\ref{C_BH_tot}) we expect that there will be
a second preasymptotic regime with $C^{(BH)}_{T_F}({\vec k},t)$ dominating
also for $x<x^*$ and producing the multiscaling behavior with
$\alpha(x)=\alpha_T(x)$ for all $x$, before standard scaling
with $\alpha(x) = 3$ is eventually established.

\section{Scalar Cahn-Hilliard-Cook model}
\label{SecCDS}
In this Section we analyze the overall time evolution of the structure factor
for the ordering dynamics of the scalar CHC.
The main point is that the dependence of the preasymptotic behavior on
the parameters of the quench $\Delta$ and $T_F$ follows patterns very similar
to those just described in the BH model.

We consider first the zero temperature quench of the $d=3$
system with small initial fluctuations $\Delta=8.3 \times 10^{-8}$.
Apart from quantitative aspects to be specified below, it is evident
from the inspection of Fig.~(\ref{Fig11}), that the basic elements in the
evolution of the structure factor are the same as in the analogous plot of
Fig.~(\ref{Fig01}) for the BH model.
Here too there is an early regime characterized by linear behavior of the peak
and rapid growth of power laws on the tails.
The early regime is abruptly followed by the onset of coarsening which,
as in the BH model, displays an intermediate regime delimited by an anomaly
in the position and the height of the peak, before the eventual late stage
is entered.
It is then interesting to see if, as in the BH model, this intermediate
regime can be associated to the manifestation of non linear mean field
behavior.
Actually, that it is so it can be established in an even more clear cut way
than in the BH model.
In fact, the mean field growth exponent is $z=4$ while in the scalar system
$z=3$.
These are precisely the values we find from $k_m(t)$ (Fig.~(\ref{Fig16}))
which gives $1/z=0.250 \pm 0.003$ in the intermediate regime and
$1/z=0.321 \pm 0.005$ in the late stage. From $C(k_m,t)$ in
Fig.~(\ref{Fig17}) we find consistently $d/z=0.730 \pm
0.002$ in the intermediate regime while no clear power law can be extracted
from the data of the late stage.
The existence of the mean field scaling regime around 
the peak in the intermediate
regime is also demonstrated by the data collapse in Fig.~(\ref{Fig12}).
The fit to the form (\ref{scalingfunction}) of the scaling function
works well for preasymptotic scaling, as in Fig.~(\ref{Fig02}), while
is inadequate for the asymptotic behavior.
Finally, the comparison of the behavior of the effective exponent
$\alpha(x,t)$ in Fig.~(\ref{Fig13}) and in Fig.~(\ref{Fig05}) completes the
evidence for the close similarity in the overall time evolution in the CHC
and BH models with the same quench parameters.
The difference in the space dimensionality is inessential.
There are, instead, important quantitative differences, as mentioned above,
in various features of the structure factor.
Thus for $k \to 0$ the scalar system has a $k^4$ tail while in the BH model one
finds the $k^2$ tail.
On the other side, for large $k$, the Porod's tail $k^{-(d+1)}$ of the scalar
case is replaced by the exponential decay in the BH model and the
analytical form of the structure factor is manifestly different around the
peak in the two models.
These differences show that certainly the BH model cannot be used for an
accurate computation of the scaling function in a scalar system, but are
immaterial when considering the gross features of the overall time evolution,
as we are doing here.
Thus, we have enough evidence to establish the sequence 
``early linear -intermediate mean field - fully non linear late stage'' 
combined with the dependence of the
crossover time on the wave vector, much in the same way as 
we have found in the BH model.
We can go one step further and regard this phenomenology,
by analogy with the BH model, as the manifestation of the 
competition among different components in the structure factor.
If so, the phenomenology should respond to the variation of the
parameters of the quench with patterns similar to those already observed in
the BH model.
As a matter of fact, taking $\Delta=1$ the overall behavior of the structure factor changes
from that of Fig.~(\ref{Fig11}) to the one in Fig.~(\ref{Fig14}), closely
reproducing the change in morphology observed between Fig.~(\ref{Fig01})
and Fig.~(\ref{Fig06}).
The same observation applies to the behaviors of $k_m$ and $C(k_m,t)$.
Comparing Figs.~(\ref{Fig16}) and~(\ref{Fig17}) with Figs.~(\ref{Fig03})
and~(\ref{Fig04}) the similarity is very close and it is definitely
established upon comparing Fig.~(\ref{Fig15}) with Fig.~(\ref{Fig08}) for
the effective exponent $\alpha(x,t)$.
Therefore, by making $\Delta$ large a considerable extension of the
intermediate regime is observed, as in the BH model, and the manifestation of
the mean field behavior takes place in the same way, with
$\alpha(x) \simeq \alpha_0(x)$ for $x<x^*$ and $\alpha(x) \simeq d=3$ for
$x>x^*$.

In the scalar case, however, a less rich variety of preasymptotic behaviors
can be obtained by varying the parameters of the quench, due to much
more rigidity in the model, in the sense that $N=1$ does not allow to modulate
at will the strength of the non linear term, which is always large.
In particular, because of the upper bound $T_F<T_c$ it is not possible
to realize the conditions for the temperature term to be overwhelmingly
larger than the non linear term.
For this reason the phenomenology observed with $T_F>0$ is not significantly
different from the one illustrated above (see Fig.~(\ref{Fig16}) and
Fig.~(\ref{Fig17})).

\section{Conclusions}
\label{SecConc}
In this paper we have studied the global evolution of the structure factor
from the very beginning of the quench down to the fully developed late stage in
a system with conserved order parameter. By making a comparative study
of the BH and CHC model  under variation of the parameters of the quench
$\Delta$ and $T_F$ we have been able to identify generic features
in the preasymptotic behavior. Prominent among these are i) the wave vector 
dependence of the time duration of the early and intermediate
regimes and ii) the mean field character of the preasymptotic
behavior. These properties are tightly linked together. The existence
of a mean field regime requires, in fact, multiscaling behavior
which in turn implies a wave vector dependence of the crossover
time. On the whole, the complex phenomenology of the preasymptotic behavior
can be accounted for in a fairly simple manner through the 
competition of different contributions as exemplified in the
structure of Eq.~(\ref{C_BH_tot}).
While for the BH model this is a direct consequence of the form of the
equation of motion, the applicability of Eq.~(\ref{C_BH_tot}) to the
evolution of the CHC model is a non trivial new result. It should be
noted that the preasymptotic phenomenology presented above is, in principle,
experimentally observable and that it would be interesting to
perform a multiscaling analysis of early time experimental data in order
to look for the crossover from mean field to truly asymptotic
behavior.

On the basis of the results produced, the BH model qualifies for
a reference theory of phase-ordering kinetics which captures
most of the qualitative ingredients entering in the phenomenology of
the process. In other words, the model can be used as the
starting point for theoretical work aiming at an improvement
of the approximation in order to have a better performance in
the computation of asymptotic properties such as the shape of
the scaling function.
In closing, it should be mentioned that the structure of the
overall time evolution analyzed in this paper is by no means
limited to the case of conserved order parameter. The crossover
from mean field to asymptotic behavior is expected to be a
relevant feature of the preasymptotic regime also when the order
parameter is not conserved. However, in the latter case it is
much more delicate to detect than in the former. With conserved
order parameter the crossover is made evident by the 
change in the growth exponent from $z=4$  to $z=3$ and, most of all, by
the spectacular change from multiscaling to standard scaling.
These elements of discrimination are absent in the non conserved
case, since the crossover does not involve any change in the value
$z=2$ of the growth exponent and scaling is standard both in the
mean field and asymptotic regimes. Work in this direction is in
progress.

\appendix
\section*{}

In this Appendix we compute the form of the effective exponent
$\alpha(x,t)$ during the linear regime for finite quench temperature
and vanishing initial fluctuations.
When $T_F>0$ and $\Delta=0$ the structure factor~(\ref{9.3}) takes three
possible forms depending on the range of the wave vectors considered.
 
\noindent If $2k^2(k^2+r)t \ll -1$ 
\be
C(k,t) = - {1 \over k^2+r} \exp \left[ -2k^2(k^2+r)t \right ].
\label{9.6}
\ee 
If $|2k^2(k^2+r)t| \ll 1$
\be
C(k,t) = 2 k^2 t.
\label{9.7}
\ee 
If $2k^2(k^2+r)t \gg 1$
\be
C(k,t) = {1 \over k^2+r}.
\label{10.1}
\ee 

The boundaries between such ranges are fixed by the conditions
\be
2 k^2 (k^2+r)t=1
\hspace{1cm}
\mbox{and}
\hspace{1cm}
2 k^2 (k^2+r)t=-1.
\ee
The first has only one real positive solution
\be
k_1(t) = \sqrt{-r + \sqrt{r^2+2/t}\over 2}
\ee
for all times.
The second has two real positive solutions
\be
k_2(t) = \sqrt{-r + \sqrt{r^2-2/t}\over 2}
\hspace{1cm}
k_3(t) = \sqrt{-r - \sqrt{r^2-2/t}\over 2}
\ee
but only for $t>2/r^2$.

Hence, two distinct linear regimes can be identified, depending on whether
the time is greater or smaller than $t_{0} \equiv 2/r^2$.
For $t < t_{0}$ the $k>0$ axis is divided into two intervals by $k_{1}(t)$.
For $k \gg k_1(t)$ the argument of the exponential in Eq.~(\ref{9.3})
is large and positive yielding the form~(\ref{10.1}),
while it is small for $k \ll k_1(t)$
and the structure factor takes the form~(\ref{9.7}).
Notice that $k_1(t)$ is proportional to $t^{-1/4}$ for small $t$.

Later on, for $t>t_{0}$, the wave vector axis is divided into four
intervals by $k_{1}(t) > k_2(t) > k_3(t) > 0$.
For $k \gg k_1(t)$, $C({\vec k},t)$  still has the form~(\ref{10.1}).
In the range $k_3(t) < k < k_2(t)$, the structure factor takes the
form ~(\ref{9.6}).
In the two other intervals ($k<k_3(t)$ and $k_2(t)<k<k_1(t)$)
it maintains the form~(\ref{9.7}). The values of
$k_3(t)$ and $k_2(t)$ coincide for $t=t_{0}$, while
they go asymptotically to zero and to $\sqrt{-r}$, respectively,
for large time.

Using Eqs.~(\ref{9.6})-(\ref{10.1}) and Eq.~(\ref{logMS}) 
the form of $\alpha(x,t)$ is readily computed.
For a structure factor obeying Eq.~(\ref{10.1}) and for large
$x$ one finds
\be
\alpha(x,t) = {2 \log {\cal L}_2(t) \over \log {\cal L}_1(t)}.
\ee
When Eq.~(\ref{9.7}) holds one has
\be
\alpha(x,t) = {\log (t/{\cal L}^2_2(t)) \over \log {\cal L}_1(t)}
\ee
and finally, when Eq.~(\ref{9.6}) is valid, neglecting logarithmic corrections
and defining $k_0=\sqrt{-r/2}$ one finds
\be
\alpha(x,t) = {2 k_0^4 t \over \log {\cal L}_1(t)} \psi(x).
\ee
Notice that only in the latter case $\alpha(x,t)$ does actually depend on $x$.

In order to fully determine $\alpha(x,t)$ the explicit time
dependence of the scaling lengths
${\cal L}_1(t) = L(k_mL)^{2/d-1}$ and ${\cal L}_2(t) = k_m^{-1}$ is needed
and this requires the determination of the time dependence of the peak
position $k_m$ during the two linear regimes.
One can easily see that, for $t\ll t_{0}$, $k_m$ decreases as $t^{-1/4}$
while for longer times $t\gg t_{0}$ it reaches the constant value
$k_m = k_0$.
Using these expressions we then have for $t \ll t_{0}$
\be
{\cal L}_1(t) = t^{1/4}\hspace{2cm}
{\cal L}_2(t) = t^{1/4}
\ee
while for $t \gg t_{0}$ 
\be
{\cal L}_1(t) = t^{1/4}\hspace{1cm}
(d=2)\hspace{2cm}
{\cal L}_1(t) = k_0^{-1/3} t^{1/6}\hspace{1cm}(d=3)
\ee
\be
{\cal L}_2(t) = k_0^{-1} = \mbox{constant.}
\ee
 
We are finally able to write down the form of the effective exponent 
in the linear regime with finite final temperature. 
For $t \ll t_{0}$
\be
\alpha(x,t) = 
\left \{
\begin{array}{rcl}
2 & \hspace{2cm} & x \ll x_1(t) \equiv k/k_1(t) \\
2 & \hspace{2cm} & x \gg x_1(t).
\end{array}
\right.
\ee
For $t \gg t_{0}$ and $d=2$
\be
\alpha(x,t) = 
\left \{
\begin{array}{lcl}
4 {\log(k_0^2 t) \over \log t}   & \hspace{2cm} &
x \ll x_2(t) \equiv k/k_2(t) \\
{8 k_0^4 t \over \log t} \psi(x) & \hspace{2cm} &
x_2(t) \ll x \ll x_3(t) \equiv k/k_3(t) \\
4 {\log(k_0^2 t) \over \log t}   & \hspace{2cm} & x_3(t) \ll x \ll x_1(t) \\
{8 \log (k_0^{-1}) \over \log t} & \hspace{2cm} & x \gg x_1(t)
\end{array}
\right.
\ee
while for $d=3$
\be
\alpha(x,t) = 
\left \{
\begin{array}{lcl}
6 {\log(k_0^2 t) \over \log (k_0^{-2} t)} & \hspace{2cm} & x \ll x_2(t) \\
{12 k_0^4 t \over \log (k_0^{-2} t)} \psi(x) & \hspace{2cm} &
x_2(t) \ll x \ll x_3(t) \\
6 {\log(k_0^2 t) \over \log (k_0^{-2} t)} & \hspace{2cm} & x_3(t)
\ll x \ll x_1(t) \\
{12 \log (k_0^{-1}) \over \log (k_0^{-2} t)} & \hspace{2cm} & x \gg x_1(t).
\end{array}
\right.
\label{12.4}
\ee

The most relevant feature of these expressions is that $\alpha(x,t)$ vanishes
for large $x$ as time grows, as opposed to the case with $\Delta=0$
where the spectrum predicted by linear theory 
becomes very large and negative for $x>\sqrt{2}$, as time grows.

\bfi
\caption{
Spectrum of multiscaling exponents in the large-$N$ model
with d=3, $T_F=0$ ($\alpha_0(x)$) and $T_F>0$ ($\alpha_T(x)$).
}
\label{Figold}
\efi
\bfi
\caption{
Structure factor in the BH model with $d=2$, $N=10^3$, 
$\Delta=10^{-8}$ and $T_F=0$.
Different curves correspond to exponentially growing times.
The same applies to all other plots of the structure factor.
}
\label{Fig01}
\efi

\bfi
\caption{
Time evolution of the inverse peak position of the structure factor in 
the BH model.
}
\label{Fig03}
\efi

\bfi
\caption{
Time evolution of the height of the structure factor peak in the BH model.
}
\label{Fig04}
\efi

\bfi
\caption{
Scaled structure factor of Fig.~(\ref{Fig01}).
}
\label{Fig02}
\efi

\bfi
\caption{
Effective exponent $\alpha(x,t)$ for short times in the BH model 
with $d=2$, $N=10^3$, $\Delta=10^{-8}$ and $T_F=0$.
The different curves are computed for subsequent non overlapping time
intervals of duration exponentially growing. The same applies to all
other plots of the multiscaling spectrum.
}
\label{Fig00}
\efi

\bfi
\caption{
Effective exponent $\alpha(x,t)$ for long times in the 
BH model with $d=2$, $N=10^3$, $\Delta=10^{-8}$ and $T_F=0$.
}
\label{Fig05}
\efi

\bfi
\caption{
Structure factor in the BH model with $d=3$, $N=10^3$, 
$\Delta=1$ and $T_F=0$. 
}
\label{Fig06}
\efi

\bfi
\caption{
Effective exponent $\alpha(x,t)$ in the BH model with $d=3$, $N=10^3$,
$\Delta=1$ and $T_F=0$.
}
\label{Fig08}
\efi

\bfi
\caption{
Effective exponent $\alpha(x,t)$ in the BH model with $d=3$, $N=10^6$,
$\Delta=0$ and $T_F=10^{-3}$.
}
\label{Fig09}
\efi

\bfi
\caption{
Effective exponent $\alpha(x,t)$ in the BH model with $d=3$, $N=10^6$,
$\Delta=1$ and $T_F=10^{-3}$.
}
\label{Fig10}
\efi

\bfi
\caption{
Structure factor in the
CHC model with $d=3$, $\Delta=8.3 \times 10^{-8}$ and $T_F=0$.
}
\label{Fig11}
\efi

\bfi
\caption{
Time evolution of the inverse peak position of the structure factor in 
the CHC model with $d=3$.
}
\label{Fig16}
\efi

\bfi
\caption{
Time evolution of the height of the structure factor peak in the
CHC model with $d=3$.
}
\label{Fig17}
\efi

\bfi
\caption{
Scaled structure factor of Fig.~(\ref{Fig11}).
}
\label{Fig12}
\efi

\bfi
\caption{
Effective exponent $\alpha(x,t)$ for long times in the CHC model with $d=3$,
$\Delta=8.3 \times 10^{-8}$ and $T_F=0$.
}
\label{Fig13}
\efi

\bfi
\caption{
Structure factor in the
CHC model with $d=3$, $\Delta=2.083$ and $T_F=0$.
}
\label{Fig14}
\efi

\bfi
\caption{
Effective exponent $\alpha(x,t)$ in the CHC model with $d=3$,
$\Delta=2.083$ and $T_F=0$.
}
\label{Fig15}
\efi

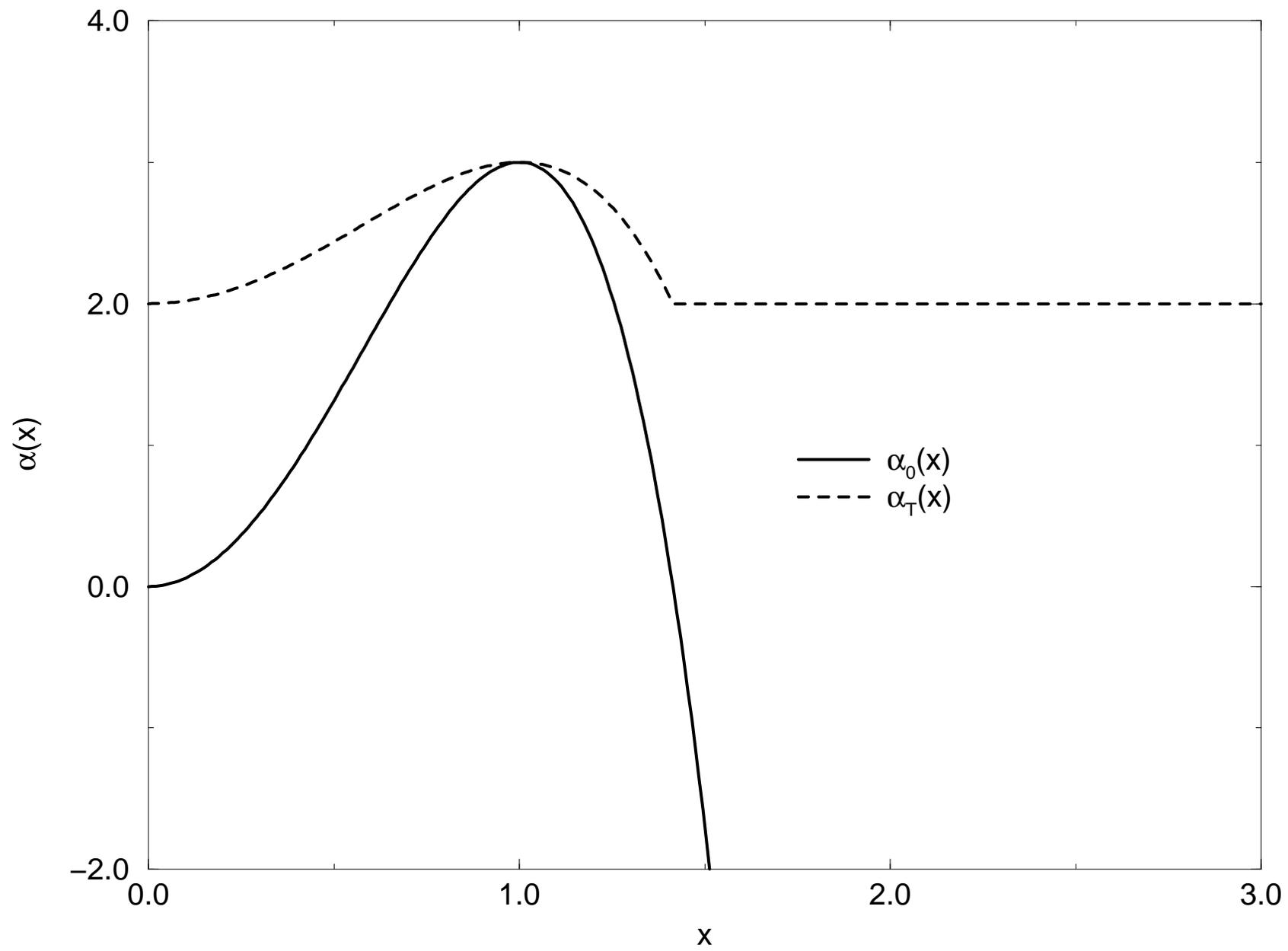
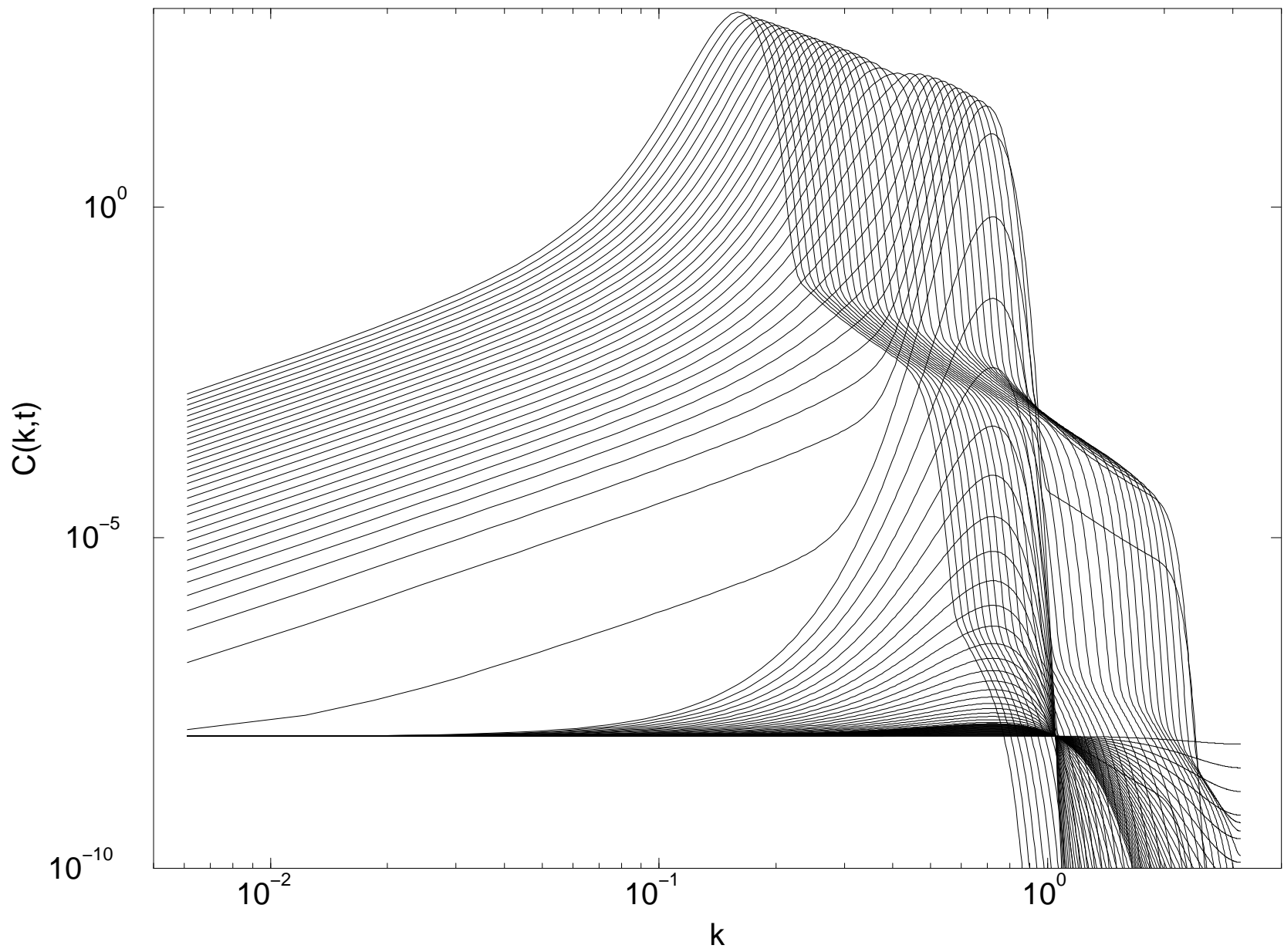
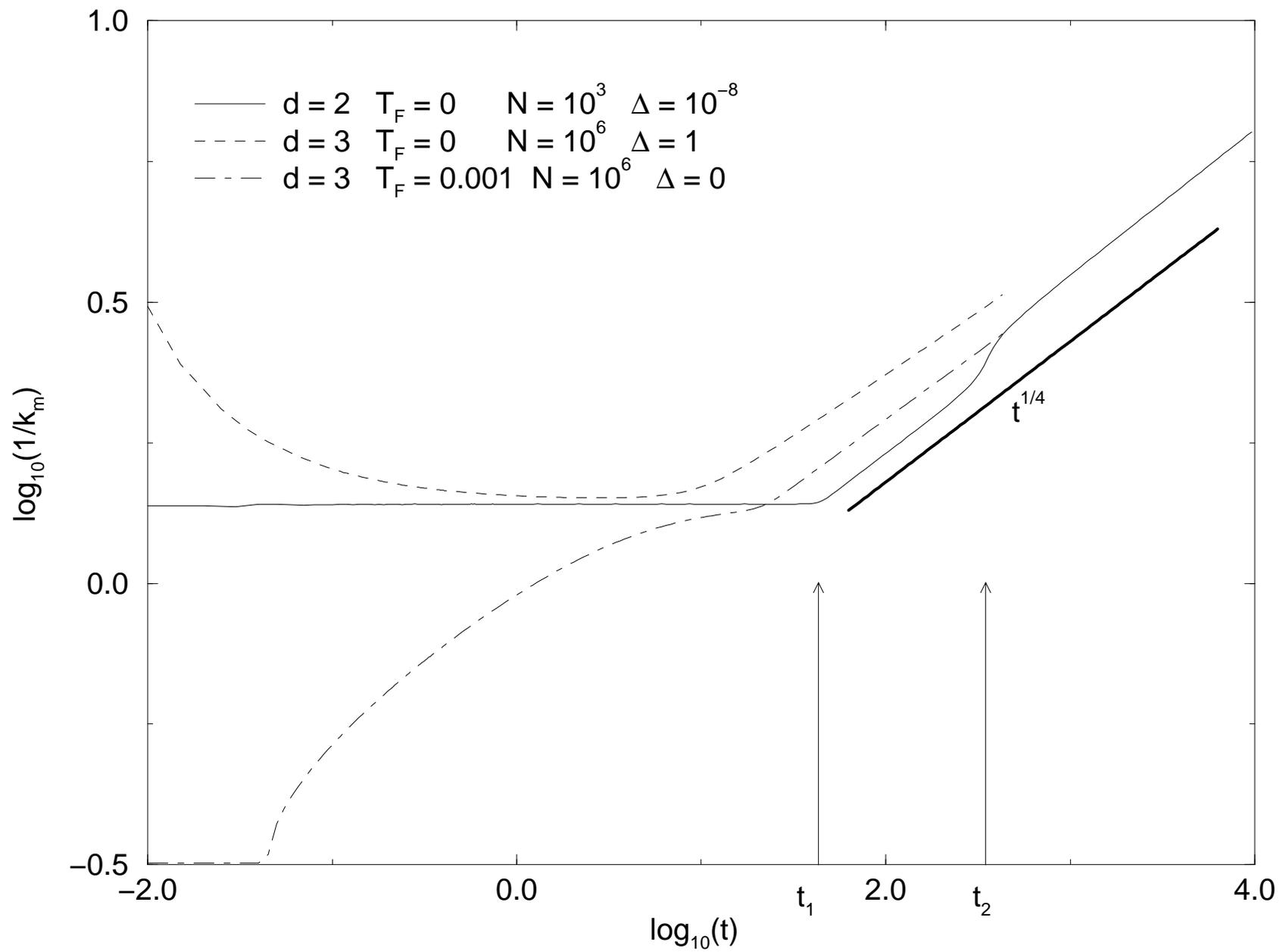
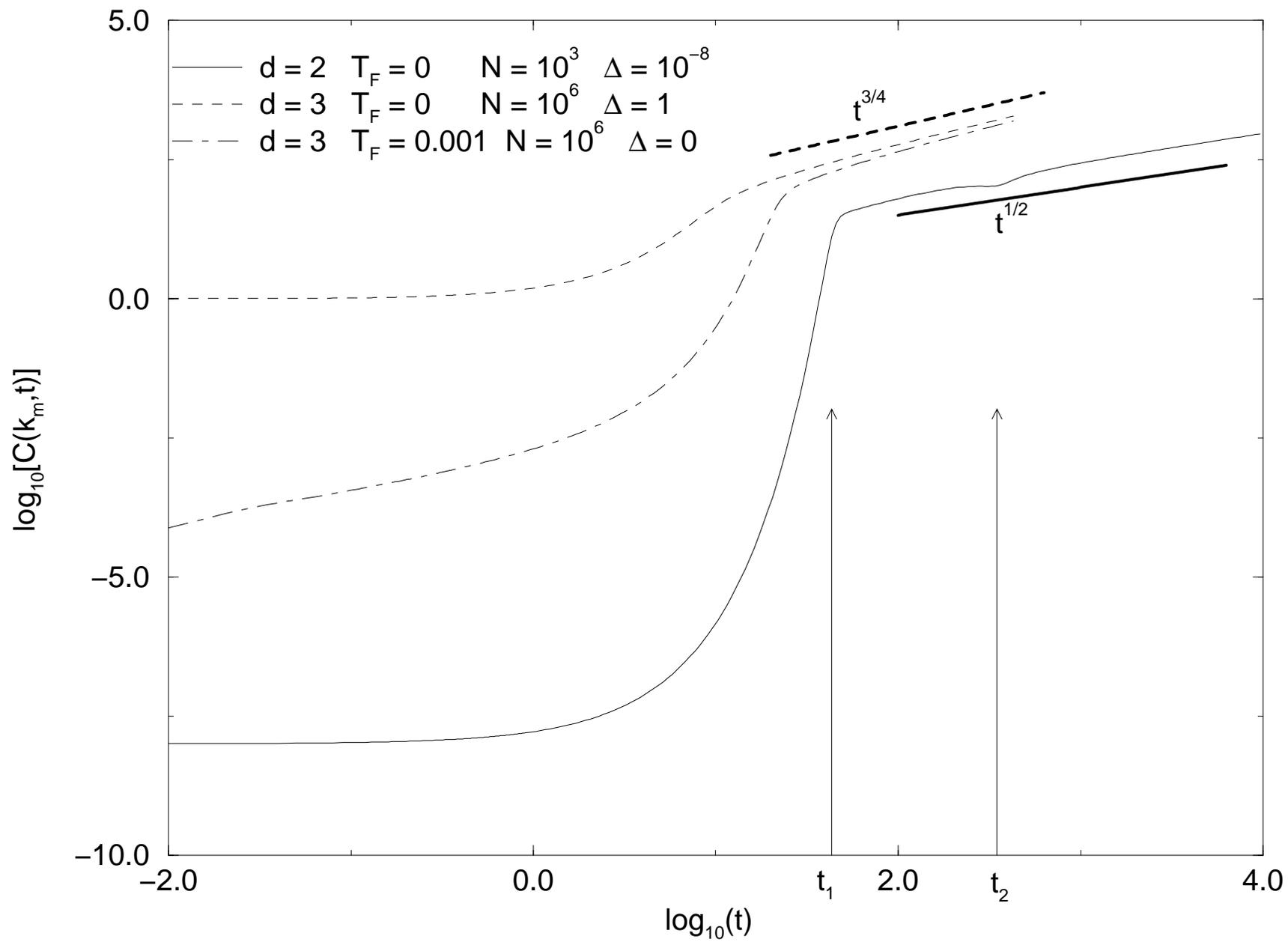
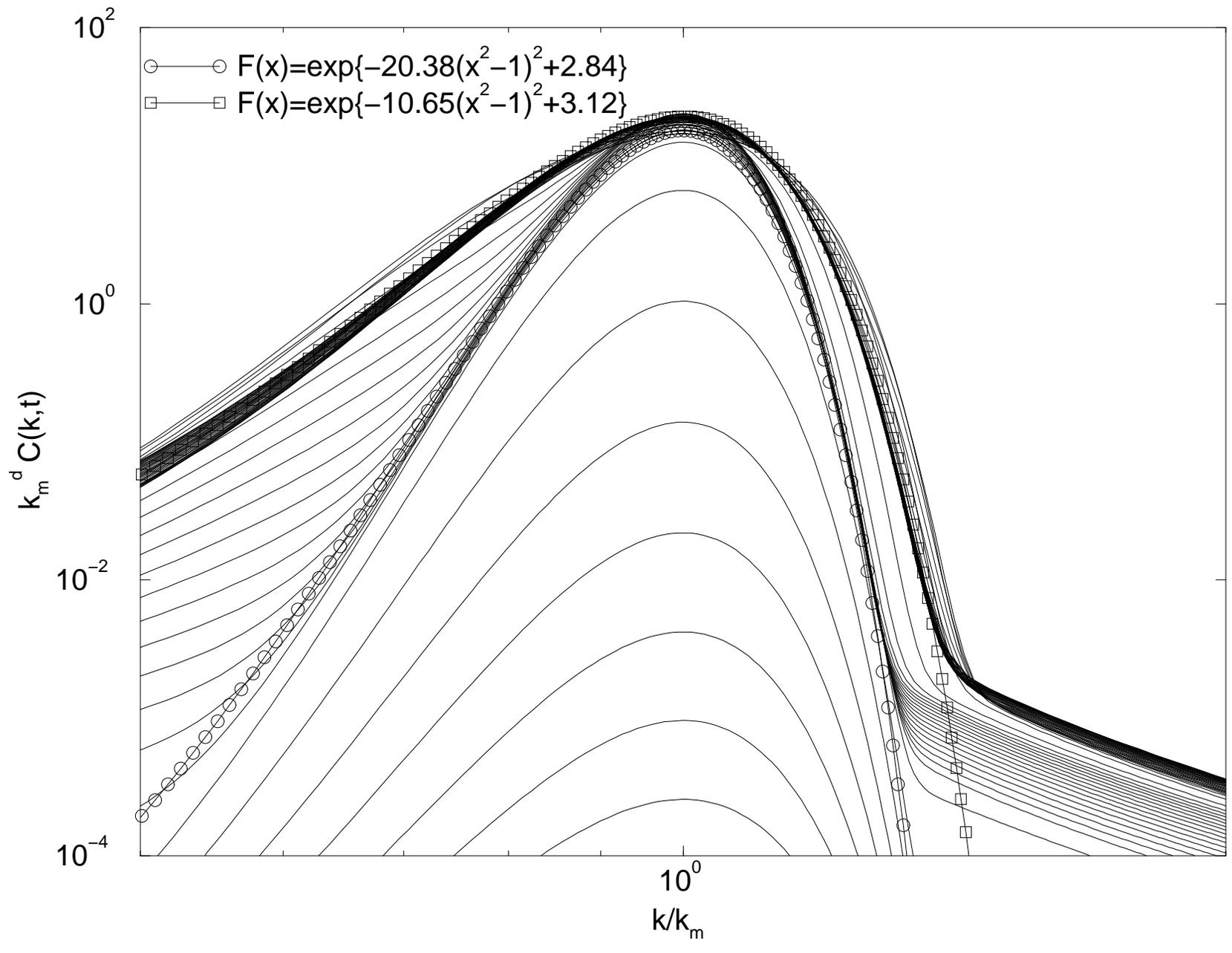
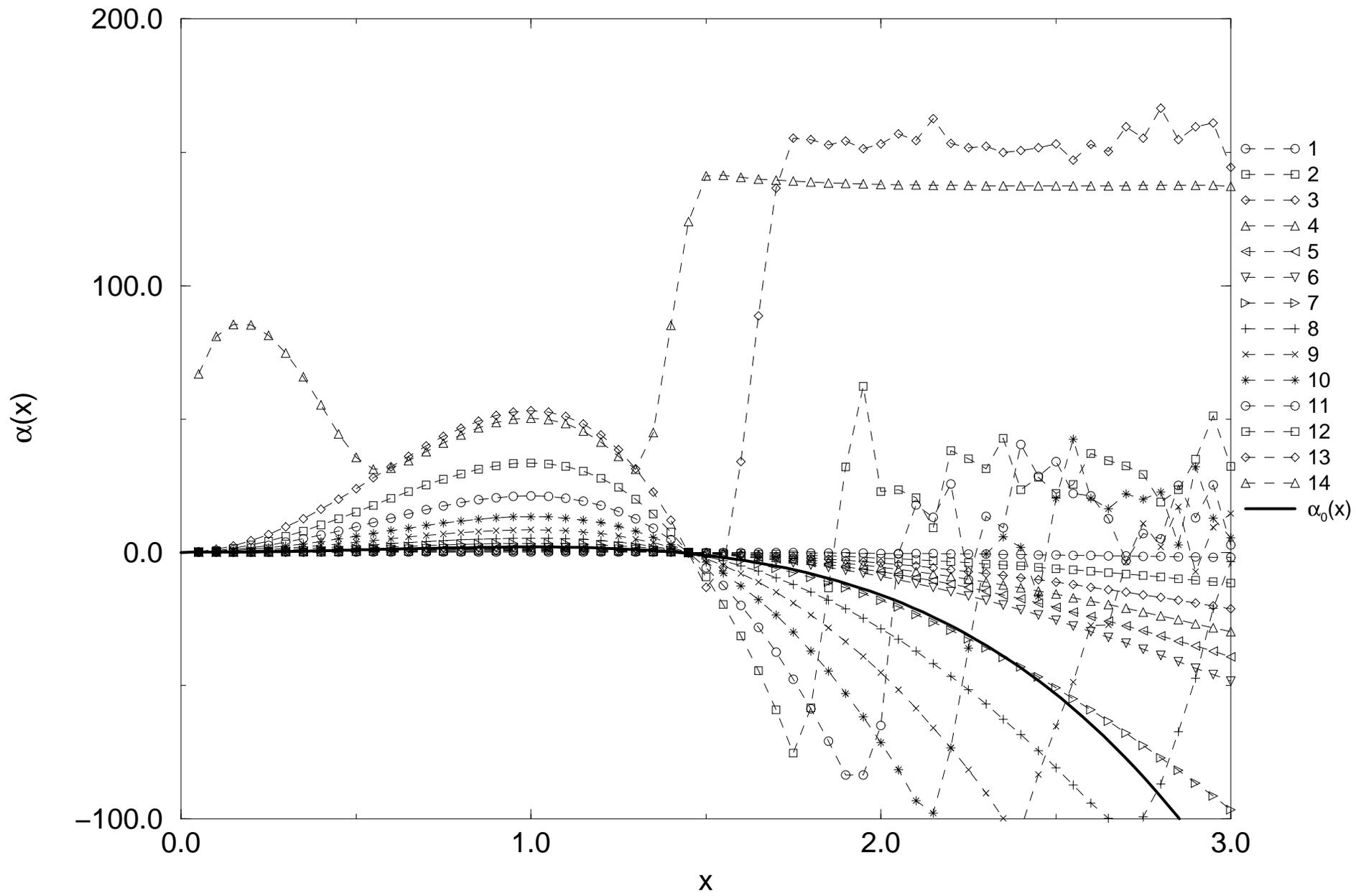
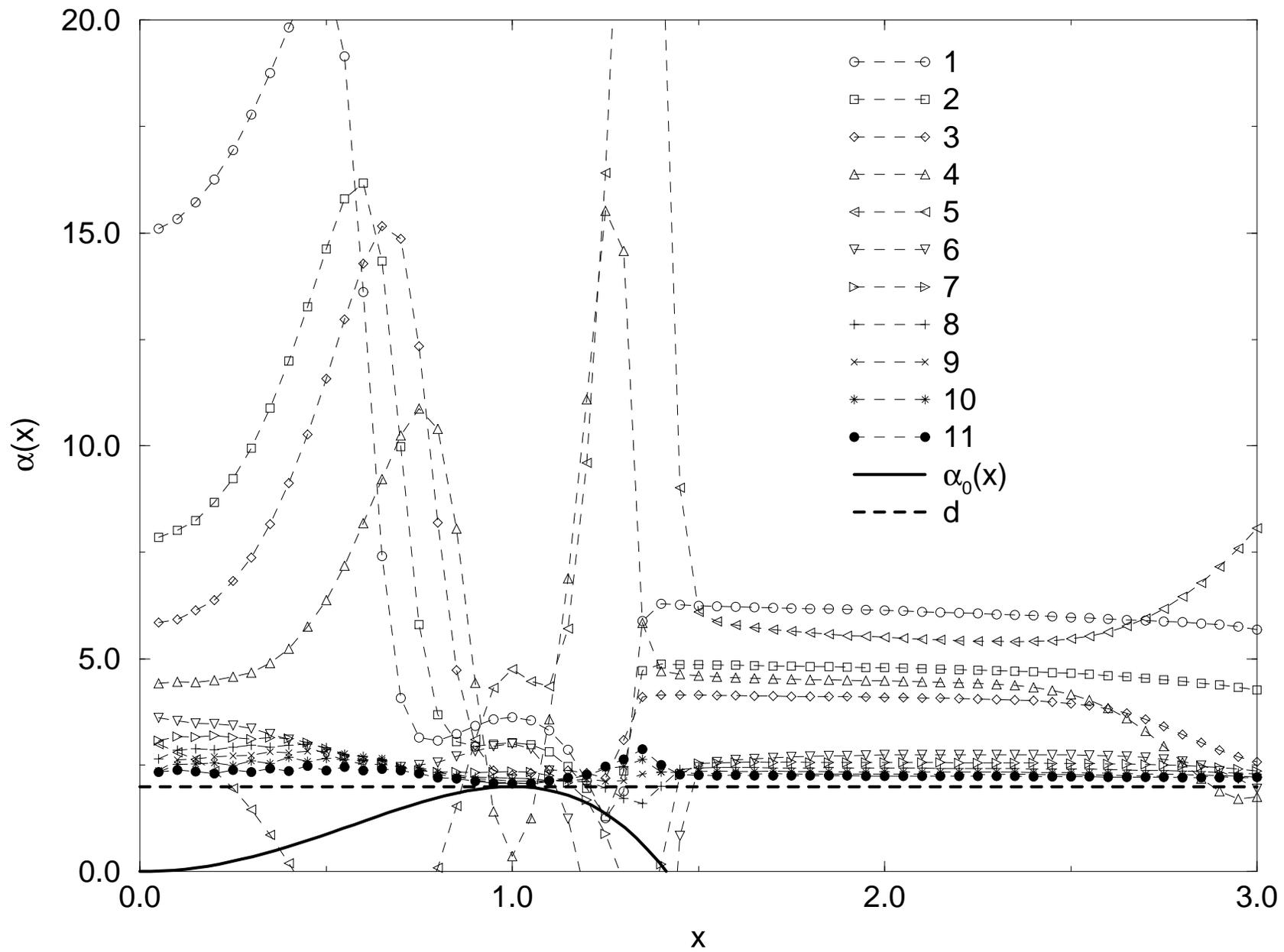
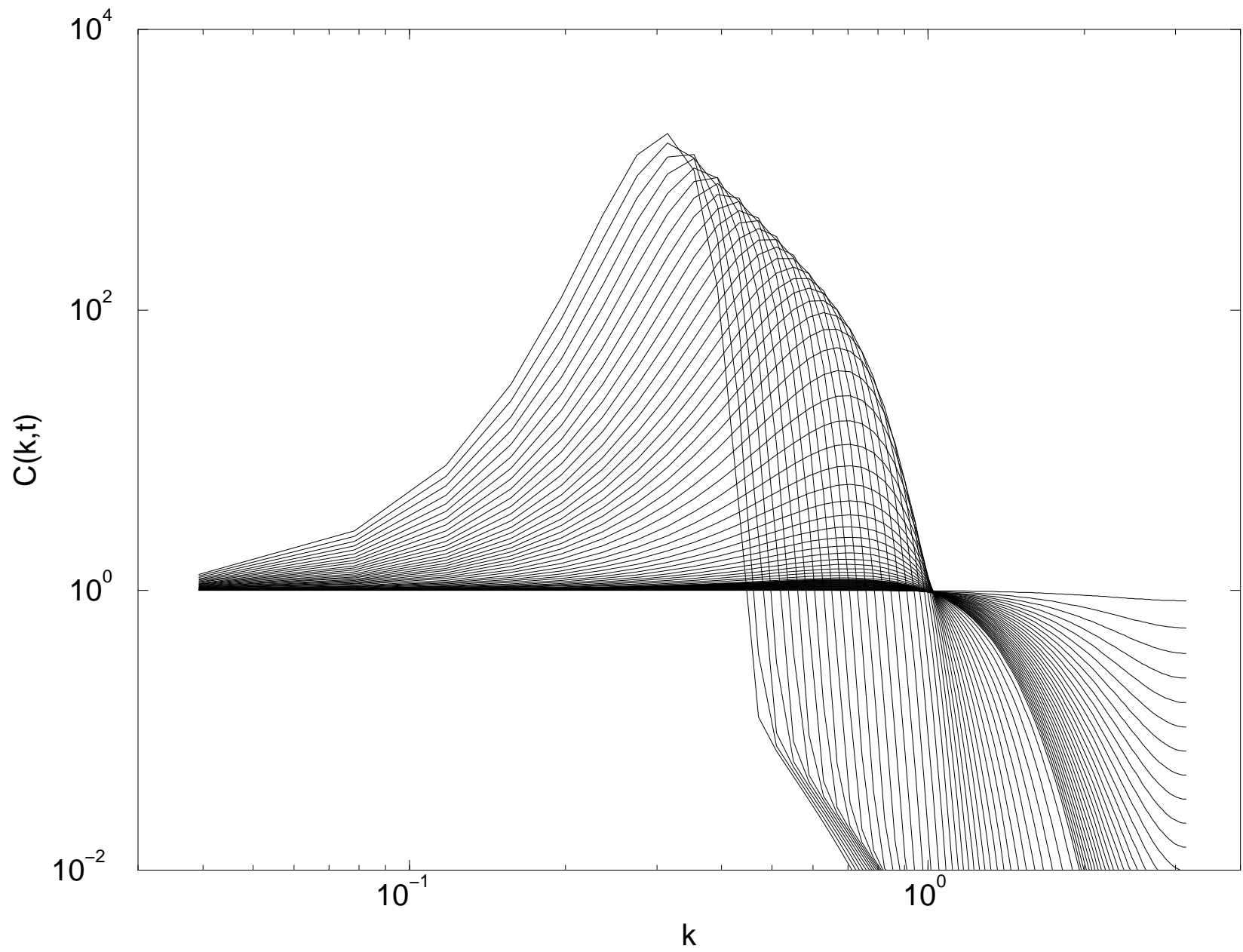
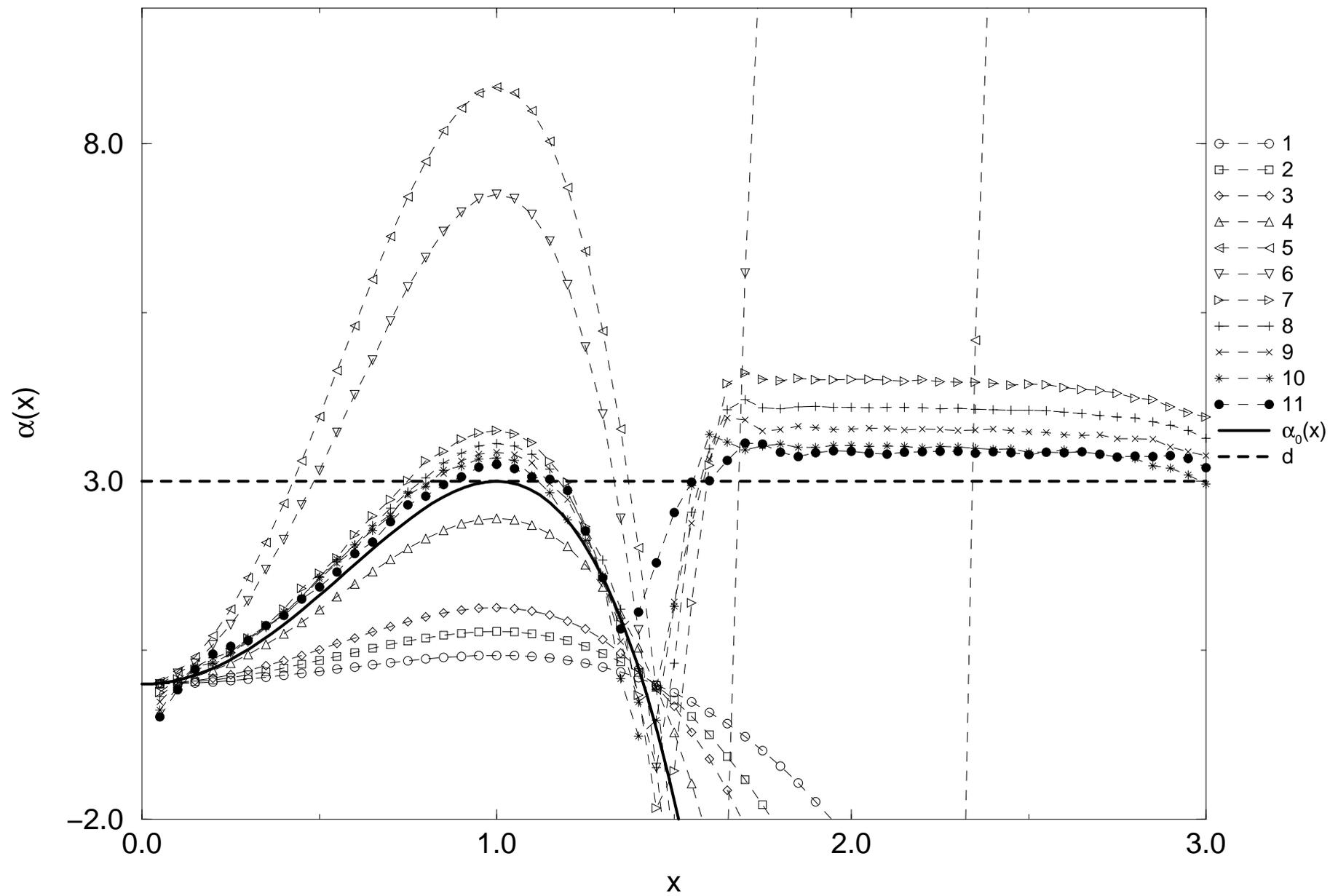
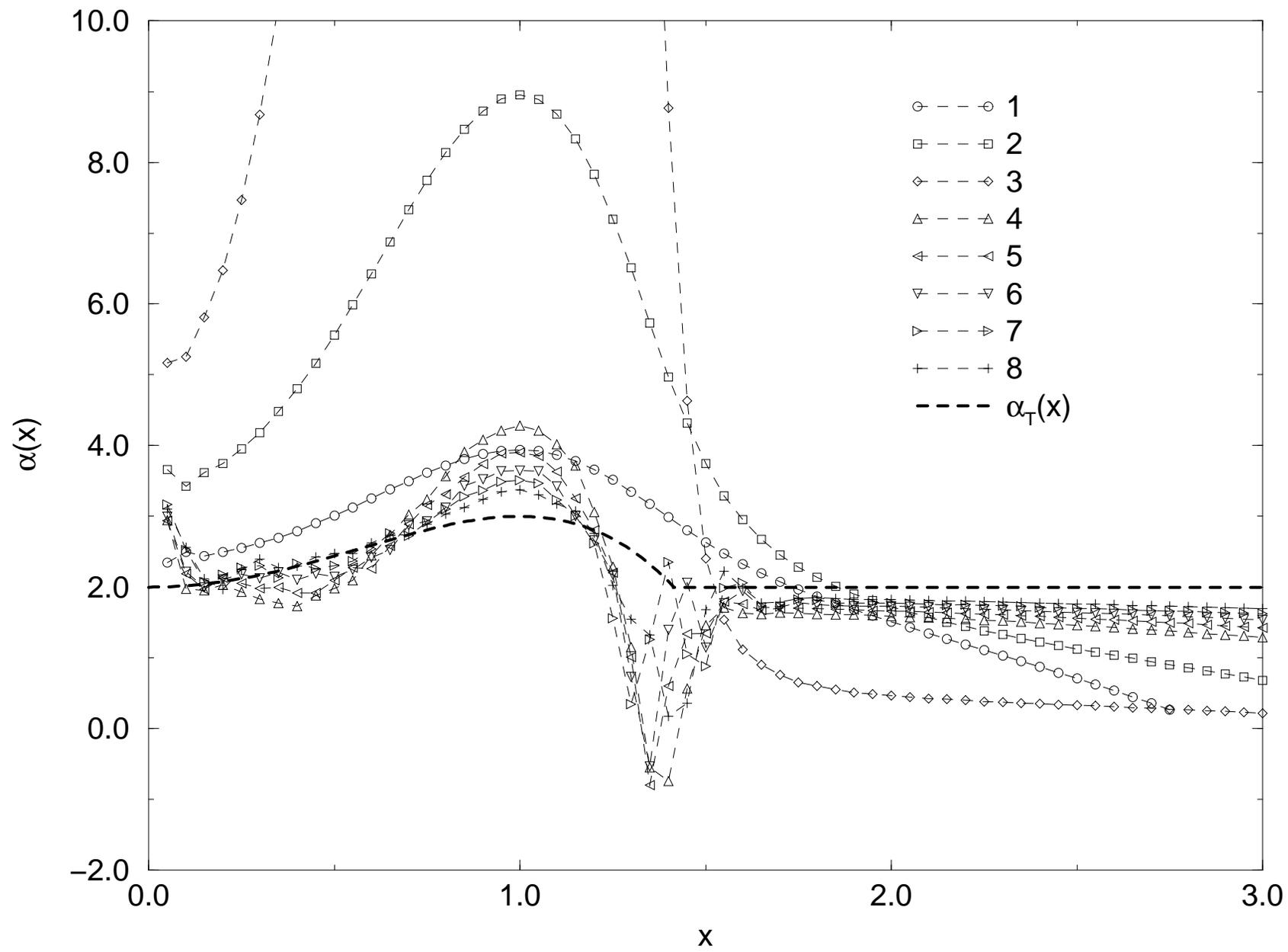
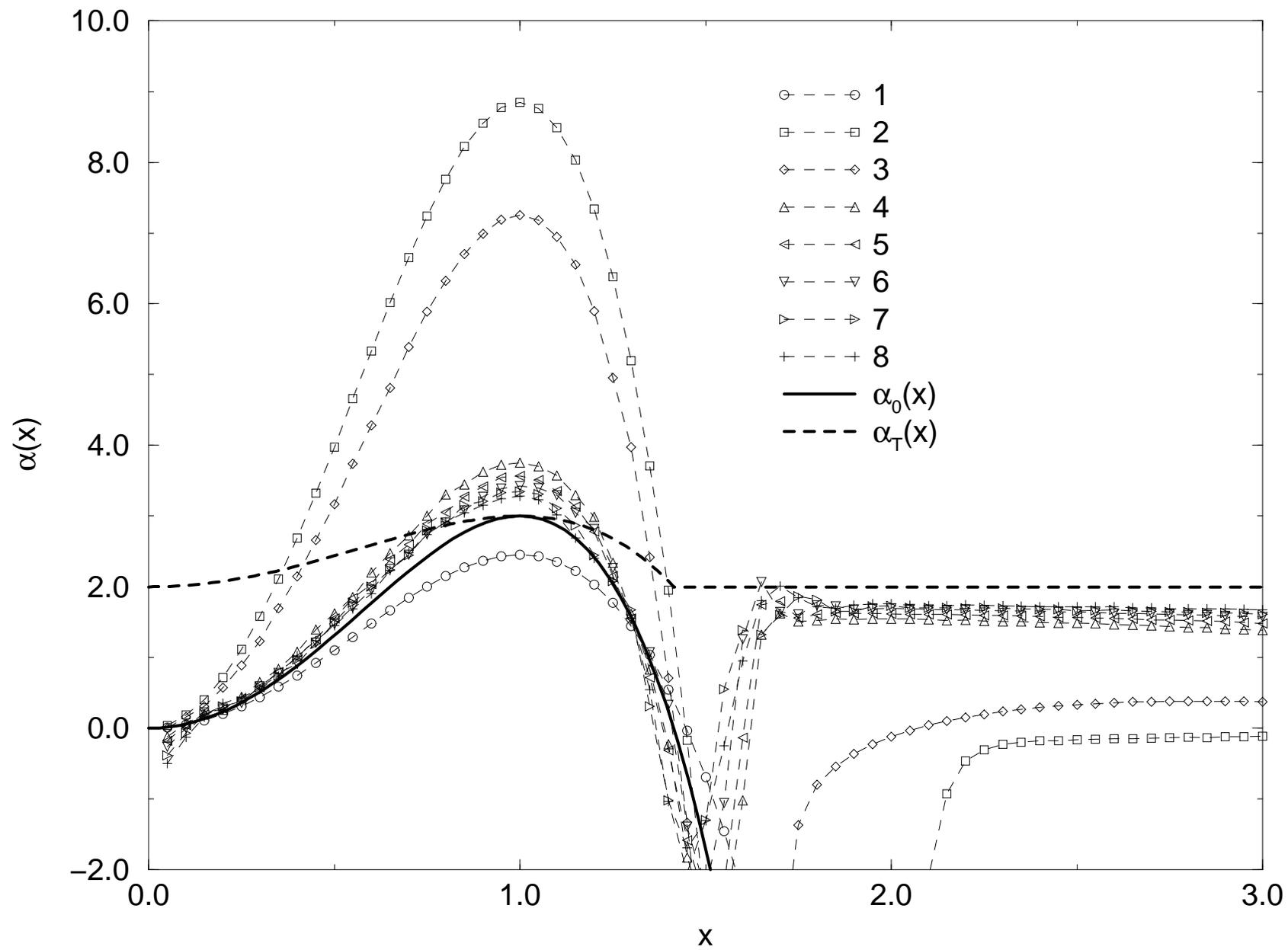
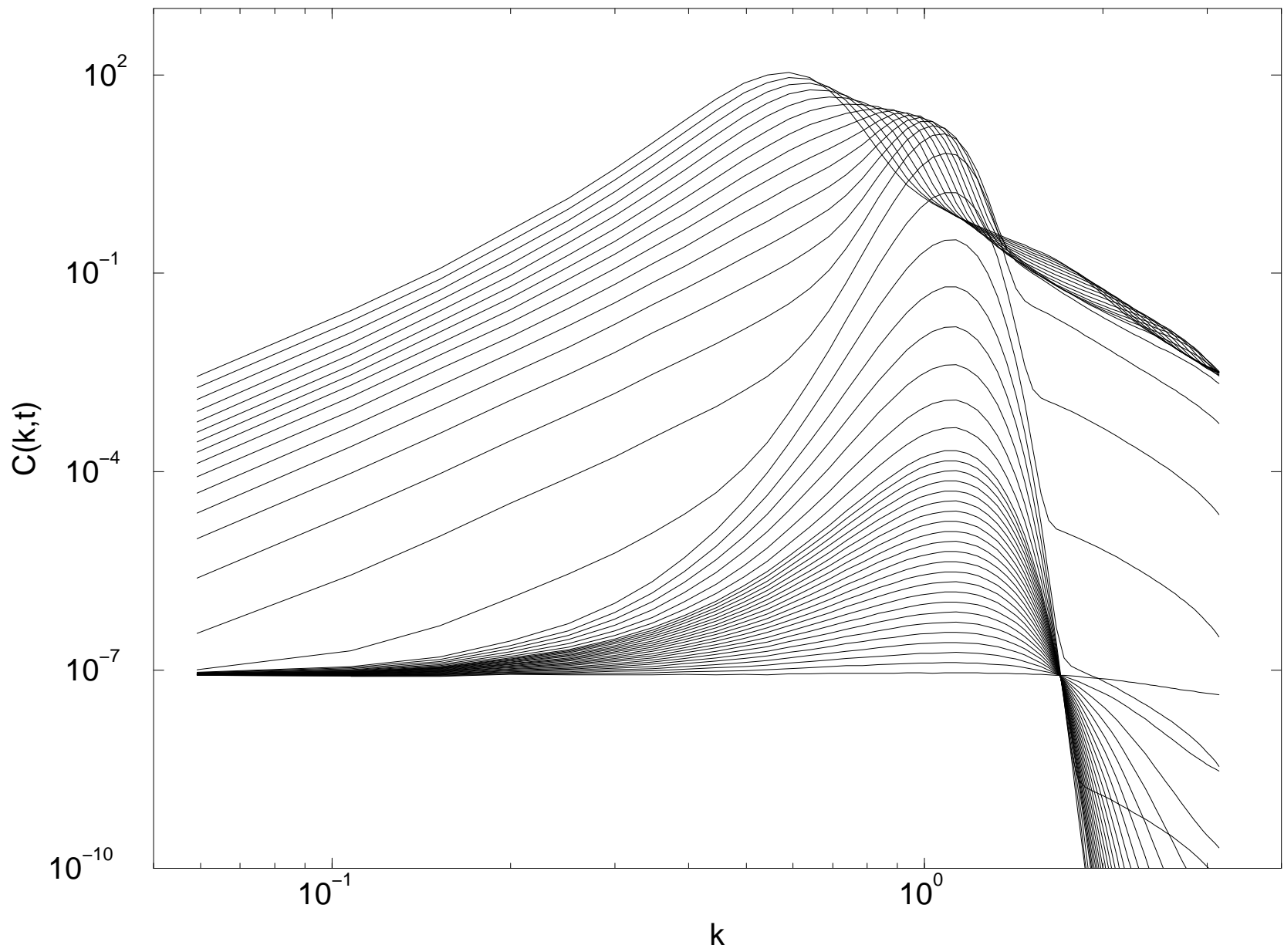
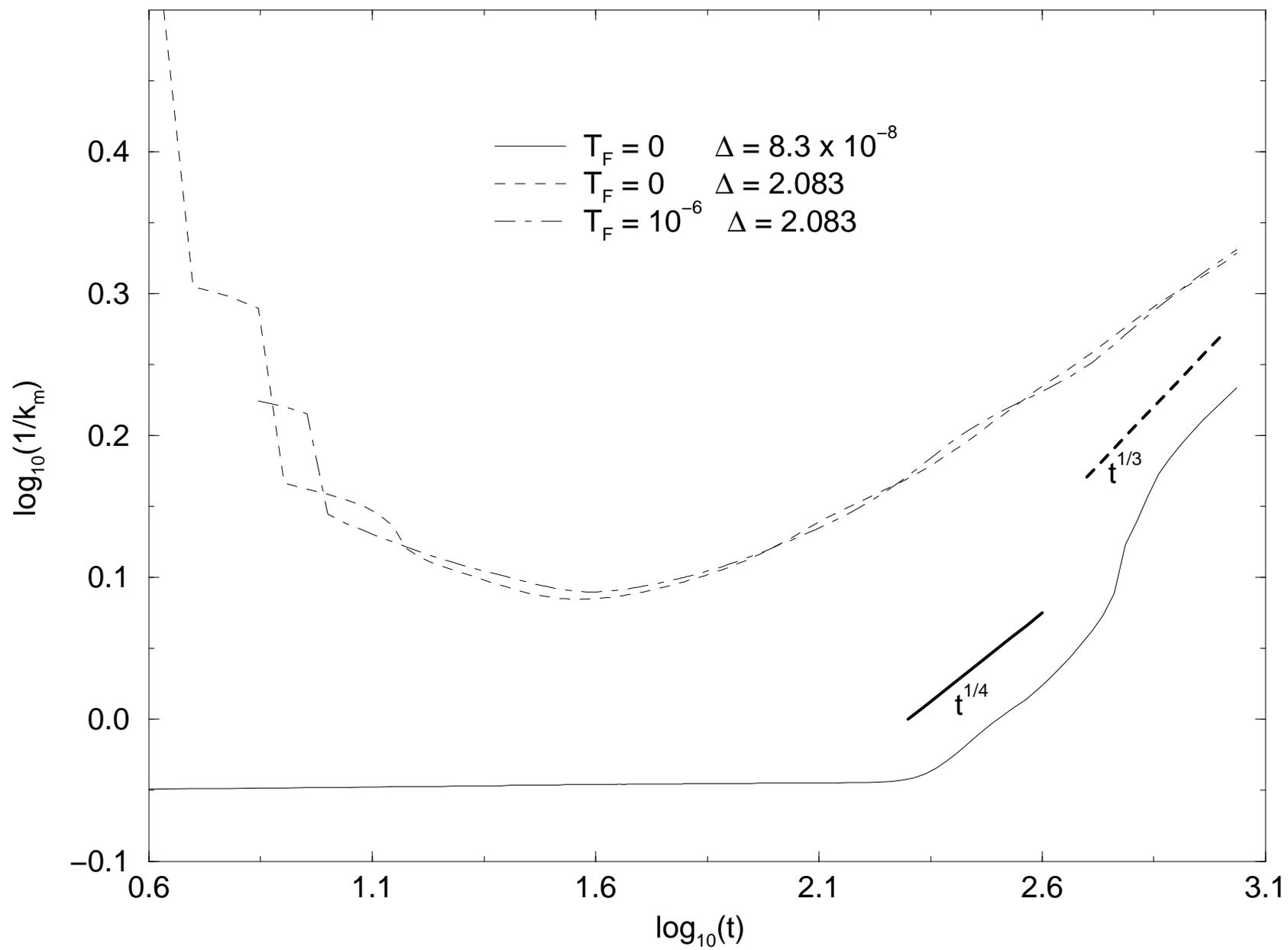
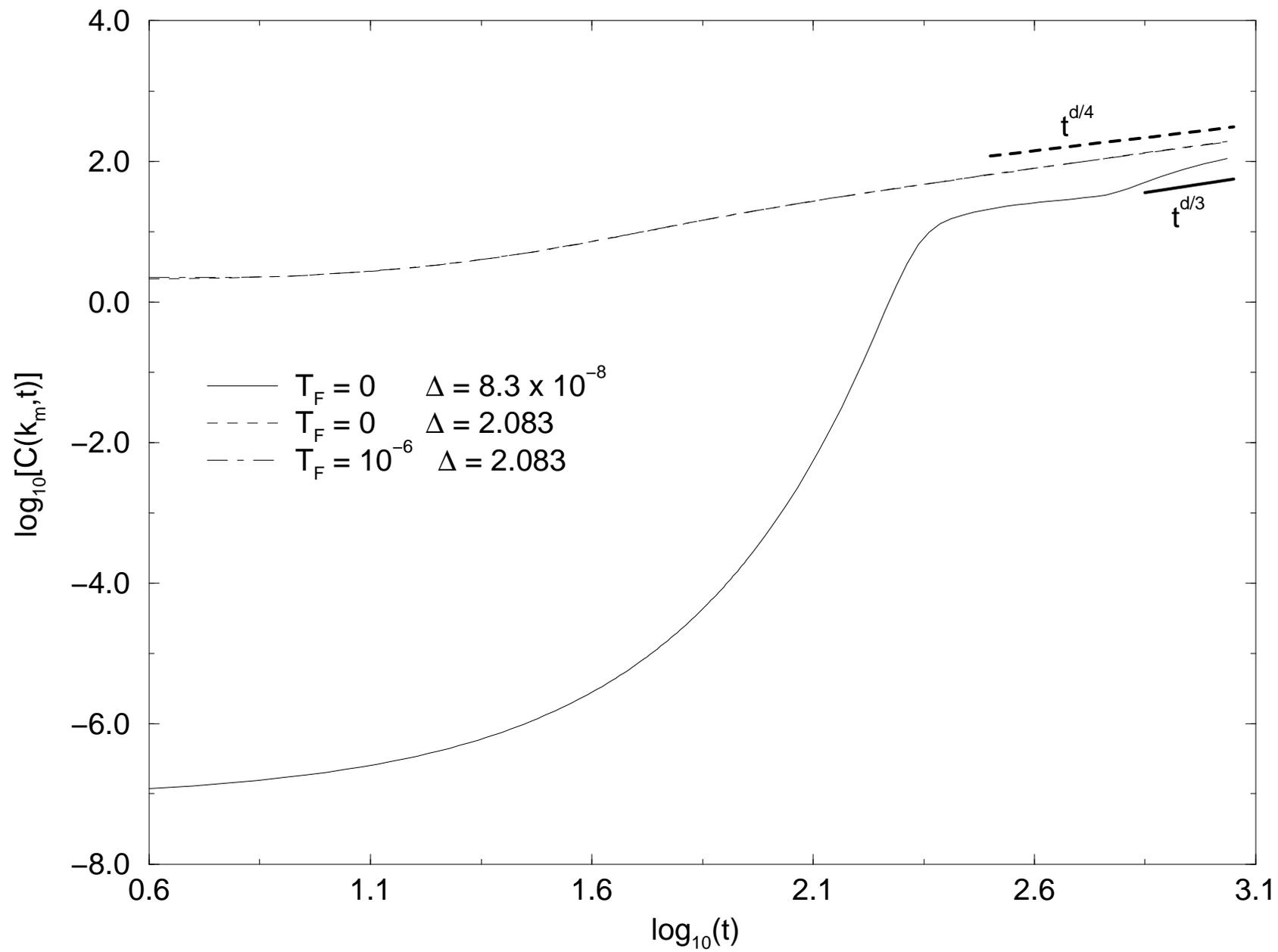
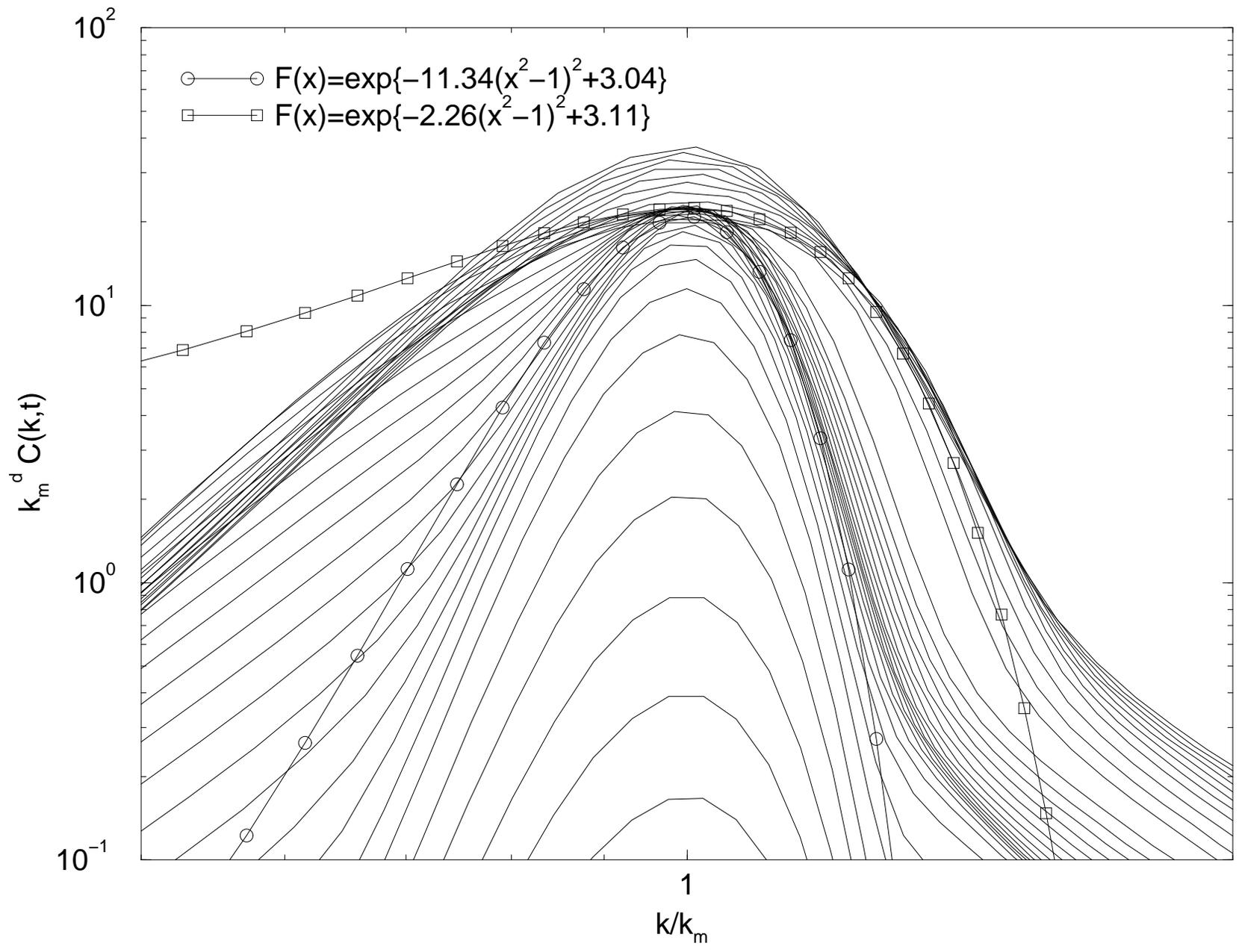
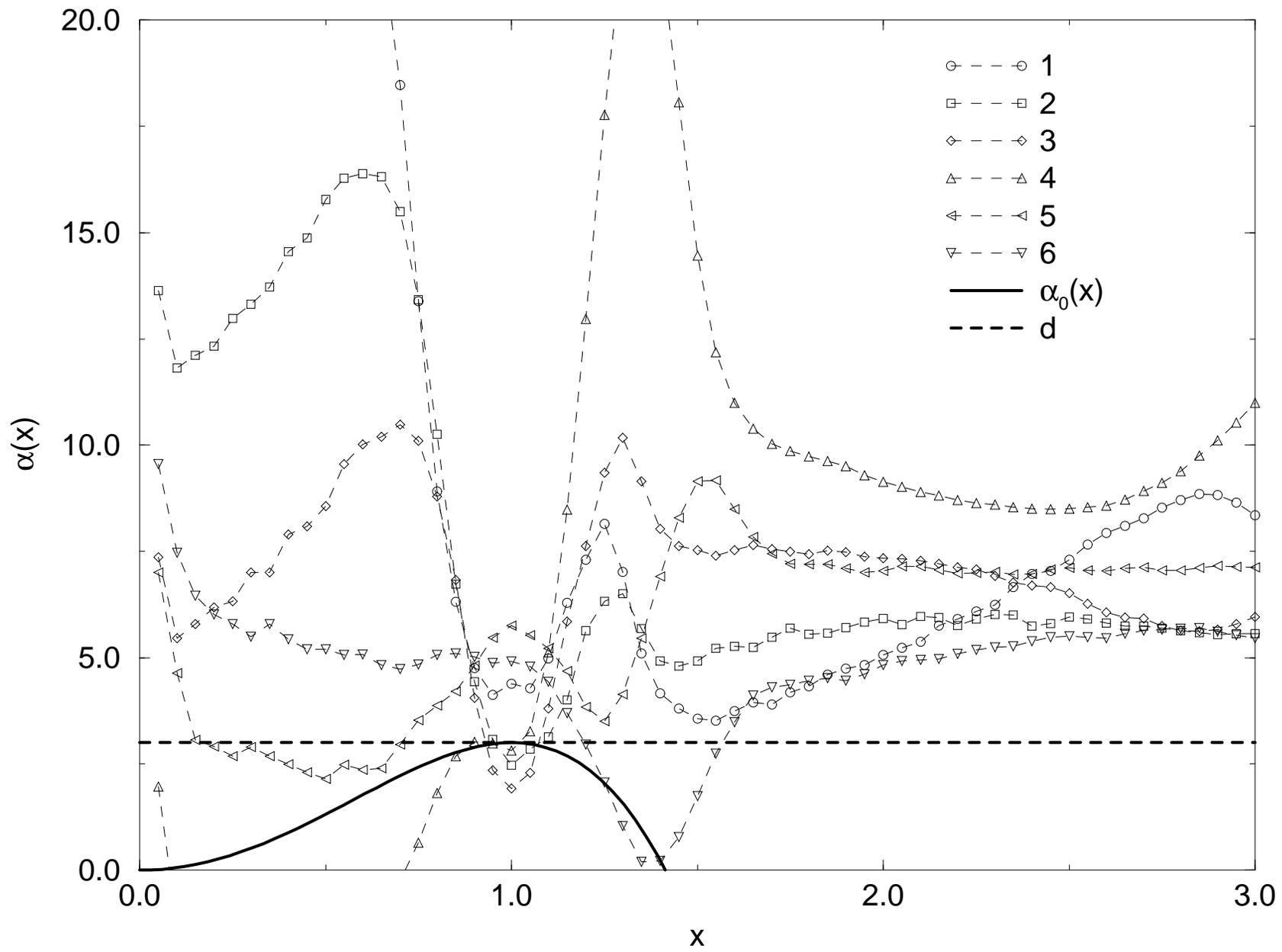
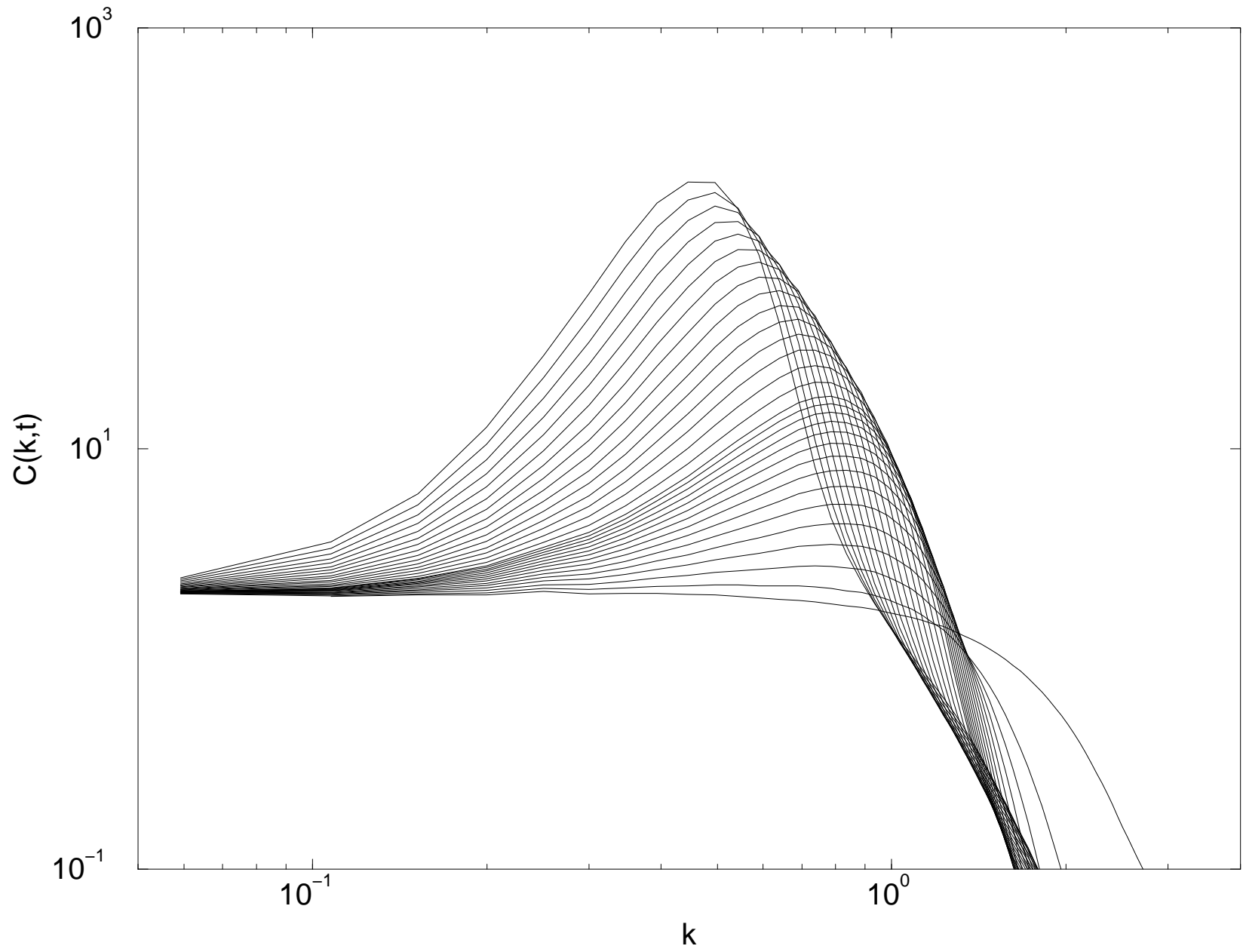
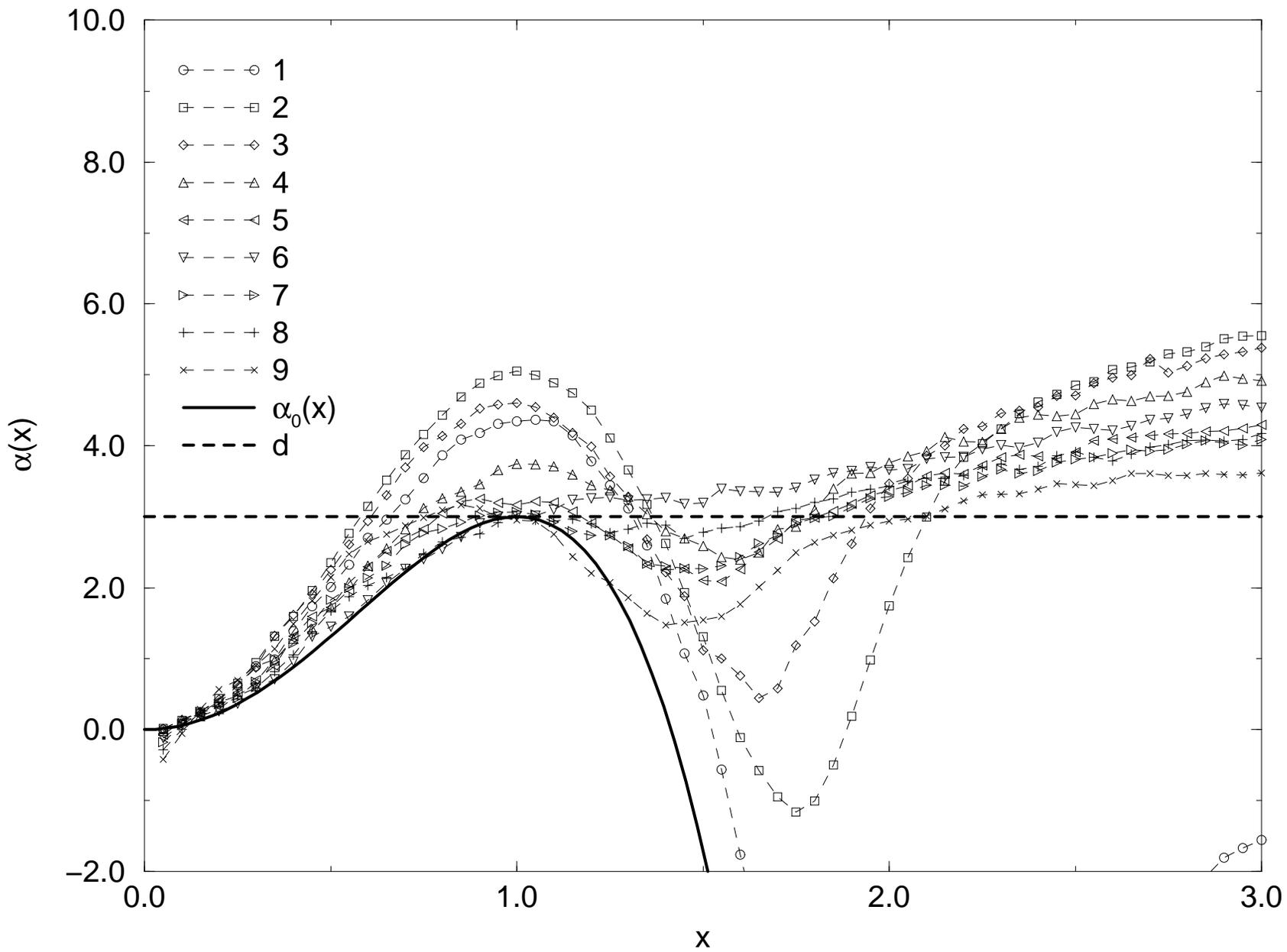
\end{document}